\documentclass[twoside]{IEEEtran} %

\usepackage[utf8]{inputenc}
\usepackage[LGR,T1]{fontenc}
\newcommand{\textgreek}[1]{\begingroup\fontencoding{LGR}\selectfont#1\endgroup}

\usepackage{amsmath,amsfonts,amssymb,mathrsfs}
\usepackage{graphicx}

\usepackage{caption}
\usepackage{subcaption}
\usepackage{color}

\usepackage{url}
\urldef{\stegano100}\url{http://csserver.ucd.ie/~fbalado/stegano100.html}

\def\argmin{\mathop{\rm arg\,min}}

\def\diag{\mathop{\rm diag}}
\def\tr{\mathop{\rm tr}}
\def\E{\mathop{\textrm{E}}}
\def\Var{\mathop{\textrm{Var}}}
\def\H{H}
\def\I{I}

\DeclareSymbolFont{bbold}{U}{bbold}{m}{n}
\DeclareSymbolFontAlphabet{\mathbbold}{bbold}
\newcommand{\ind}[1]{\mathbbold{1}_{\{#1\}}}

\title{Asymptotically Optimum Perfect Universal\\ Steganography of
  Finite Memoryless Sources}

\author{Félix Balado,~\IEEEmembership{Member,~IEEE} and David
  Haughton\thanks{F.~Balado is with the School of Computer Science,
    University College Dublin, Belfield Campus, Dublin 4, Ireland
    (e-mail: felix@ucd.ie). David Haughton is with the Centre for
    Applied Data Analytics Research, Belfield Office Park,
     Dublin 4, Ireland (e-mail:
    david.haughton@ucd.ie). This research was conducted with
    the financial support of Science Foundation Ireland under Grant
    Number 09/RFP/CMS2212. Preliminary parts of this work were
    presented at the IEEE ICASSP Conference (May 2013), IEEE WIFS
    Conference (November 2013), and the WIO Conference (July 2014). The algorithms in this paper are included in the Matlab
    toolbox \textgreek{Stegan'o} 100\% available at \stegano100.

    Copyright (c) 2017 IEEE. Personal use of this material is permitted.  However, permission to use this material for any other purposes must be obtained from the IEEE by sending a request to pubs-permissions@ieee.org.
  }}

\begin{document}
\maketitle

\begin{abstract}
  A solution to the problem of asymptotically optimum perfect
  universal steganography of finite memoryless sources with a passive
  warden is provided, which is then extended to contemplate a
  distortion constraint. The solution rests on the fact that Slepian's
  Variant~I permutation coding implements first-order perfect
  universal steganography of finite host signals with optimum
  embedding rate. The duality between perfect universal steganography
  with asymptotically optimum embedding rate and lossless universal
  source coding with asymptotically optimum compression rate is
  evinced in practice by showing that permutation coding can be
  implemented by means of adaptive arithmetic coding. Next, a
  distortion constraint between the host signal and the
  information-carrying signal is considered. Such a constraint is
  essential whenever real-world host signals with memory (e.g.,
  images, audio, or video) are decorrelated to conform to the
  memoryless assumption. The constrained version of the problem
  requires trading off embedding rate and distortion. Partitioned
  permutation coding is shown to be a practical way to implement this
  trade-off, performing close to an unattainable upper bound on the
  rate-distortion function of the problem.

\end{abstract} %


\IEEEpeerreviewmaketitle

\section{Introduction}
\label{sec:introduction}
\IEEEPARstart{D}{igital} data hiding refers to coding techniques which
aim at embedding information within digital discrete-time host
signals~\cite{moulin05:_data_hiding}. In short, steganography is a
special data hiding scenario in which undetectability of the embedded
information is paramount ---unobtrusiveness, rather than
undetectability, suffices in general data hiding. In the
steganographic problem, a ``man in the middle'' (warden) performs a
detection test on signals sent between two parties in order to
determine whether they carry hidden information or not. In the
scenario considered here, the warden does not alter the tested
signals, which thus arrive unmodified at the decoder (passive
warden). Besides circumventing detection by the warden, the encoder
also wishes to maximize the steganographic embedding rate, which is
the amount of bits per host element that can be conveyed to the
decoder through the covert channel created by modifying the host in
order to embed (hide) information.

An important landmark in steganography research was the realization of
the existence of an inextricable connection between steganography and
source coding. Anderson and Petitcolas gave an intuitive rationale for
this link in the early days of digital steganography~\cite[Section
VI-A]{anderson98:_limits}. These authors pointed out that if we would
have lossless source coding with optimum compression rate for
real-world signals (i.e., ideal compression for signals such as
digital images), then these signals would have to be dense in the
space of the optimum source code. Thus, decompressing any arbitrary
sequence from this space would always render a true real-world
signal. In other words, a lossless compression algorithm with optimum
compression rate could also work as a perfect (undetectable)
steganographic algorithm with optimum embedding rate, by using
decompression to encode a message into an information-carrying signal
and compression to decode that message.

This duality between perfect steganography with a passive warden and
lossless source coding %
implied that practical steganographic algorithms had to be intimately
related to practical compression algorithms. Soon, some authors
partially succeeded in translating this fundamental relationship into
well-founded steganographic methods. The first proposal along these
lines was in the early work of Cachin~\cite{cachin98:inform}. Later
on, Sallee added another important piece to the puzzle with
model-based steganography~\cite{sallee03:model}. Subsequent research
has gradually drifted away from these seminal contributions, and
steganography has become a subject for the most part disconnected from
source coding. Machine learning has grown in importance, and relevant
information-theoretical results such as~\cite{comesana07}
or~\cite{wang08:_perfec_secur} have been largely
sidelined. %

Here we present a contribution which we believe fills an important gap
in the field: the asymptotically optimum solution to the canonical
steganography problem dual of universal lossless compression of
memoryless signals with optimum compression rate. Furthermore, we
consider the implications of the application of this solution to real-world
signals such as multimedia, and we show its connections with existing
results about steganographic systems. The roots of the questions
considered here are found in Cachin's
criterion~\cite{cachin98:inform}. This criterion tells us that a
perfect steganographic system is implemented by an encoder that
exactly preserves the distribution of the host signal, because in this
way optimum detection by the warden will be foiled. The implementation
of Cachin's criterion raises a crucial issue: %
what should the encoder do if the distribution of the host signal is
not known? As noted by Cachin himself by drawing the parallel with
source coding, \textit{universal} steganography must necessarily be
the matter-of-fact approach to implementing perfect steganography: it
is the empirical distribution of the host that should be
preserved~\cite{cachin98:inform}. Thus, a canonical problem in
steganography %
is how to undertake perfect universal steganography of memoryless host
signals with optimum embedding rate.  The core element of the solution
presented here is Slepian's Variant~I permutation
coding~\cite{slepian65:permutation}. The centrality of permutations to
the problem at hand was already discovered ---either explicitly or
not--- by a number of researchers over the years, most prominently by
Ryabko and Ryabko~\cite{ryabko09:_asymp_optim}. However, the answer to
the fundamental question is still open: how does one implement a general
perfect universal steganographic algorithm for finite memoryless host
signals with asymptotically optimum embedding rate?  

Cachin's criterion implicitly assumes that a probabilistic model can
completely capture the nature of the signals produced by a
steganographic encoder. However this is not generally true whenever
real-world signals meaningful to humans (e.g., images, audio, video)
are used as hosts, as the semantics of such signals are not accurately
captured by any known probabilistic model (in particular by an
empirical model on which the universal approach must be based). In the
context of this paper semantics become relevant whenever a reversible
decorrelating transform is applied to real-world signals with memory,
for them to conform to the memoryless
assumption~\cite{verdu11:_teaching}. Due to the aforementioned
limitations of the models, even if an information-carrying signal can
be produced which has the exact statistics of some model, it may still be
semantically wrong ---and thus suspicious to a human warden. By
continuity arguments, enforcing a similarity constraint between host
and information-carrying signal ---in addition to the empirical
statistics preservation constraint--- can help preserve the semantics
of the host in the information-carrying signal. This constraint
implies a second open question: what is the optimum embedding rate for
a given similarity constraint (embedding distortion constraint)?

In this paper we address the two questions outlined above. The
material is organized as follows. Section~\ref{sec:notation-framework}
introduces the notational conventions and the basic setting assumed
throughout the paper. Section~\ref{sec:prior-work} reviews prior work
on steganography of memoryless hosts and motivates our
study. Section~\ref{sec:permutation-codes} introduces the use of
permutation coding for steganography and discusses a low-complexity
implementation, which solves the first
question. Section~\ref{sec:embedd-dist-geom} is devoted to a
theoretical analysis of the embedding distortion of permutation coding
in a steganographic context. Next, Section~\ref{sec:embedd-dist-contr}
addresses the issue of embedding distortion control and describes a
suboptimal solution to the second question.  Theoretical and
empirical results are compared in Section~\ref{sec:results}, and,
lastly, Section~\ref{sec:conclusions} draws the conclusions of this
work.

\section{Preliminaries} 
\label{sec:notation-framework}
\paragraph{Notation} Boldface lowercase Roman letters are column vectors.  The $i$-th
element of vector $\mathbf{a}$ is $a_i$, or~$(\mathbf{a})_i$ whenever
this notation is more convenient. %
The special symbols~$\mathbf{1}$ and~$\mathbf{0}$ are the all-ones
vector and the null vector, respectively. Capital Greek letters denote
matrices; the entry at row $i$ and column $j$ of matrix $\Pi$ is
$(\Pi)_{i,j}$. In keeping with standard notation, the only
exception to this convention is the exchange
matrix~$\mathrm{J}$. $\mathrm{I}$ is the identity matrix. $\tr \Pi$ is the trace
of~$\Pi$. $(\cdot)^t$ is the transpose
operator. $\diag(\mathbf{a})$ is a diagonal matrix with $\mathbf{a}$ in its diagonal. The indicator function is defined
as~$\ind{\theta}=1$ if logical expression~$\theta$ is true, and zero
otherwise. The 2-norm of a vector~$\mathbf{r}$ is
$\|\mathbf{r}\|=\sqrt{\mathbf{r}^t\mathbf{r}}$. The Hamming distance
between two $n$-vectors $\mathbf{r}$ and $\mathbf{s}$ is
$\delta(\mathbf{r},\mathbf{s})=\sum_{i=1}^n\ind{r_i\neq s_i}$. The
Hamming weight of $\mathbf{r}$ is
$\omega(\mathbf{r})=\delta(\mathbf{r},\mathbf{0})$. Calligraphic
letters are sets; $|\mathcal{V}|$ is the cardinality of
set~$\mathcal{V}$. When describing algorithms, $x\leftarrow v$ means
the assignment of value $v$ to variable $x$.

A host sequence is denoted by the discrete-valued $n$-vector
$\mathbf{x}=[x_1,x_2,\ldots,x_n]^t\in\mathcal{V}^n$ where
$\mathcal{V}=\{v_1,v_2,\ldots,v_q\}\subset \mathbb{Z}$. We assume that
$\mathbf{x}\neq \mathbf{0}$, and that
$\mathbf{v}=[v_1,v_2,\ldots,v_q]^t$ gives the elements of
$\mathcal{V}$ in increasing order, that is, $v_1<v_2<\cdots<v_q$. The
histogram of~$\mathbf{x}$ is a vector $\mathbf{h}=[h_1, h_2,\ldots,
h_q]^t$ such that $h_k=\sum_{i=1}^n \ind{v_k=x_i}$ for
$k=1,2,\ldots,q$, and therefore $\mathbf{h}^t\mathbf{1}=n$; $\mathbf{v}$ is
hence the vector containing the ordered histogram bins. An
information-carrying sequence is denoted by
$\mathbf{y}=[y_1,y_2,\ldots,y_n]^t$.

Let $\mathcal{S}_n$ be the symmetric group, namely, the group of all
permutations of $\{1,2,\ldots,n\}$. We denote a permutation
$\boldsymbol\sigma\in\mathcal{S}_n$ by means of a vector
$\boldsymbol\sigma=[\sigma_1,\sigma_2,\ldots,\sigma_n]^t$ where
$\sigma_i\in\{1,2,\ldots,n\}$ and $\sigma_i\neq\sigma_j$ for all
$i\neq j$. This vector defines in turn a permutation
matrix ${\Pi}_{\boldsymbol\sigma}$ with entries
$(\Pi_{\boldsymbol\sigma})_{i,j}=\ind{\sigma_i=j}$. The reordering
of $\mathbf{x}$ using $\boldsymbol\sigma$ is the vector
$\mathbf{y}={\Pi}_{\boldsymbol\sigma}\,\mathbf{x}$, for which
$y_i=x_{\sigma_i}$ for $i=1,2,\ldots,n$. Two or more
different permutations may lead to the same reordering of the elements
of $\mathbf{x}$. For this reason we will follow the convention that a
\textit{rearrangement} of $\mathbf{x}$ is a unique ordering of its
elements. A special case is the rearrangement of $\mathbf{x}$ in
nondecreasing order. This is obtained by means of a permutation
${\boldsymbol\sigma}^\uparrow$ yielding
${\mathbf{x}^\uparrow}=\Pi_{{\boldsymbol\sigma}^\uparrow}\,\mathbf{x}$
such that ${x}^\uparrow_{\!1}\le{x}^\uparrow_{\!2}
\le\cdots \le {x}^\uparrow_{\!n}$.  The rearrangement of
$\mathbf{x}$ in nonincreasing order can be obtained from
${\mathbf{x}^\uparrow}$ as
${\mathbf{x}^\downarrow}=\mathrm{J}{\mathbf{x}^\uparrow}$,
where $\mathrm{J}$ is the exchange matrix ---the permutation matrix
with entries $(\mathrm{J})_{i,j}=\ind{j=n-i+1}$.

Italicized Roman or Greek capital letters represent random
variables. The probability mass function (pmf) of a random variable
$X$ with support $\mathcal{V}$ is denoted by $p(X=v)$, with
$v\in\mathcal{V}$, or simply by $p(v)$ if clear from the context.  We
will also refer to $\mathbf{p}=[p(X=v_1),p(X=v_2),\ldots,p(X=v_q)]^t$
as the pmf of~$X$, and to $\mathbf{v}$ as its support. The probability
of an event~$\vartheta$ is denoted by~$\Pr\{\vartheta\}$. The
expectation, variance, and entropy of $X$ are denoted by $\E\{X\}$,
$\Var\{X\}$, and $\H(X)$, respectively. The binary entropy function is
denoted by~$h(p)$. $\I(X;Y)$ is the mutual information between $X$
and~$Y$. Logarithms are base~2 throughout the paper, unless explicitly
noted otherwise. Asymptotic equalities and inequalities (as
$n\to\infty$) are marked with a dot on top of the usual sign.

\begin{figure}[t!]
  \centering
  \includegraphics[width=8.5cm]{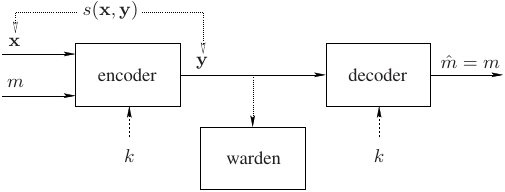}
  \caption{Setting studied in the paper: the encoder modifies a host
    signal~$\mathbf{x}$ to produce a signal $\mathbf{y}$ carrying
    message~$m$. Encoder and decoder may share a symmetric secret
    key~$k$. Further parameters are shared in the special case in
    which a constraint on the embedding distortion
    $s(\mathbf{x},\mathbf{y})$ is enforced (see
    Section~\ref{sec:embedd-dist-contr}).}

\label{fig:general}
\end{figure}

\paragraph{Setting} The setting studied in this paper is shown in
Figure~\ref{fig:general}. The encoder is a function
$e(\cdot,\cdot):\mathcal{V}^n\times\{1,2,\ldots,r\}\to \mathcal{V}^n$
which produces an information-carrying signal
$\mathbf{y}^{(m)}=e(\mathbf{x},m)$, where $\mathbf{x}$ is the host
and~$m\in\{1,2,\ldots,r\}$ is the message to be hidden. An alternative
view of the encoder is seeing it as adding a \textit{watermark}
$\mathbf{w}^{(m)}\triangleq \mathbf{y}^{(m)}-\mathbf{x}$ to the
host $\mathbf{x}$ when it wishes to embed message~$m$ in it. In the
remainder we will drop the superindex~$(m)$ whenever there is no
ambiguity, for clarity of exposition.  The decoder is a function
$d(\cdot):\mathcal{V}^n\to\{1,2,\ldots, r\}$ that retrieves the
message hidden in a received vector.  The decoded message can be put
as $\hat{m}=d\left(\mathbf{y}\right)$, and with a passive warden
$d\left(\mathbf{y}^{(m)}\right)=m$. The embedding rate (transmission
rate) is defined as $\rho\triangleq (1/n)\log r$ (bits/host
element). The closeness or similarity between $\mathbf{x}$ and
$\mathbf{y}$ is gauged through a function
$s(\cdot,\cdot):\mathcal{V}^n\times\mathcal{V}^n\to [0,\infty)$ such
as those discussed in Sections~\ref{sec:squar-eucl-dist}
and~\ref{sec:average-degree-host}; $s(\mathbf{x},\mathbf{y}^{(m)})$ can
be seen as measuring the embedding distortion caused by hiding
message $m$ in~$\mathbf{x}$.

We will assume that~$\mathbf{x}$ is drawn from a discrete memoryless
source. As discussed in Section~\ref{sec:prior-work}, achieving
perfect steganography in these conditions requires that $\mathbf{y}$
and $\mathbf{x}$ always have identical empirical distribution
(histogram). The fundamental goal is the maximization of the embedding
rate~$\rho$ under this constraint. As we have mentioned, we will also
study the maximization of~$\rho$ under a constraint on
$s(\mathbf{x},\mathbf{y})$.

As shown in Figure~\ref{fig:general}, the encoding and decoding
functions can also depend on a symmetric secret key for privacy, that
is, $e_k(\cdot,\cdot)$ and $d_k(\cdot)$. For simplicity, and without
loss of generality, we will omit this key from most of our exposition,
although we will show how to implement keyed encoding/decoding.

\section{Prior Work and Problem History}
\label{sec:prior-work}
Cachin was the first to sketch an answer to the problem of maximum
rate steganography of memoryless signals, relying on a description of
a generic universal compressor based on the method of
types~\cite{cachin98:inform}.  The construct proposed by Cachin
suffers from two shortcomings: 1) it does not provide perfect
steganography for finite hosts, as it only achieves perfection
asymptotically when the size of the host goes to infinity; 2) it
assumes that all signals with the same empirical distribution as the
host are valid outputs of the encoder.  The first shortcoming was addressed by
Ryabko and Ryabko~\cite{ryabko09:_asymp_optim}, who described a
universal steganographic algorithm with optimum embedding rate for
finite hosts. Apart from not providing a completely general
implementation, the authors of~\cite{ryabko09:_asymp_optim} do not
address the second shortcoming ---the most acute in practical
scenarios as we argue next. Recalling the duality argument by Anderson
and Petitcolas in the introduction, observe that, in their idealized
setting, the encoder produces an information-carrying signal
\textit{ab initio}, relying on an ideal model of the signals that the
encoder can output and without the need for a host signal. Yet, in a
universal approach, such as in~\cite{cachin98:inform} or
in~\cite{ryabko09:_asymp_optim}, the role of the hypothetical ideal
host model is played by the empirical model of a given host. A key
observation to be made is: nothing guarantees that all signals which
preserve the empirical model of the host will also be ``close'' to
it. This is critical when decorrelation of real-world signals with memory is
used to conform to the memoryless assumption~\cite{verdu11:_teaching}:
not all signals that preserve the first-order statistics of a host
signal in the decorrelated domain will map back to semantically
meaningful real-world signals in the original domain. As we have discussed,
this issue can be remediated by enforcing a similarity constraint
between host and information-carrying signal (embedding distortion
constraint). This constraint implies that practical steganography must
be a problem of coding with noncausal side information at
the encoder~\cite{Costa83}, where the host~$\mathbf{x}$ constitutes
the deterministic side information.

In regard to the embedding distortion constraint, a practical approach
to the problem of near-perfect steganography of finite memoryless
sources with asymptotically optimum embedding rate was given by
Sallee~\cite{sallee03:model}. This author used the quantized block
discrete cosine transform (DCT) domain as a rough approximation to a
domain where the host is memoryless, and in which theoretical
probabilistic models are available (the generalized Cauchy
distribution is used in~\cite{sallee03:model}). Exploiting these two
properties Sallee proposed a scheme called model-based steganography which
uses arithmetic coding to achieve asymptotically optimum embedding
rate while preserving a first-order model of the host, simultaneously constraining the embedding distortion. However, its
reliance on a theoretical model of the host means that model-based
steganography is neither universal nor perfect. %

For a memoryless host signal, perfect universal steganography
is achieved by preserving its first-order statistics.
If the host is finite and discrete-valued, preserving its first-order
statistics is equivalent to preserving its histogram.  A number of
previous authors more or less explicitly claim that their
steganographic algorithms implement histogram preservation. However,
on close examination, many allegedly histogram-preserving methods are
only approximations. %
We will briefly review next the few methods that do implement exact
histogram preservation, which therefore implement perfect
steganography of finite memoryless sources. Among them we find Provos'
OutGuess~\cite{provos01:_defending} ---the first histogram-preserving
steganographic algorithm---, Franz's proposal~\cite{franz03:_stegano},
Ryabko and Ryabko's method~\cite{ryabko09:_asymp_optim}, Kumar and
Newman's J3~\cite{kumar10:_j3}, and Luo and Subbalakshmi's
method~\cite{luo11:_zero_kullb_liebl}.

Most of these methods
(see~\cite{provos01:_defending,franz03:_stegano,kumar10:_j3}) are
variations of least-significant bit (LSB) steganography.  In this
early heuristic steganographic method the encoder embeds a message
into a host $\mathbf{x}$ by producing a signal $\mathbf{y}$ whose
elements are $y_i=2\lfloor x_i/2\rfloor+b_i$ for $i=1,2,\ldots,n$,
where $b_i\in\{0,1\}$ is the message bit embedded in the $i$-th
element of the host. %
The histogram-preserving methods
in~\cite{provos01:_defending,franz03:_stegano} and~\cite{kumar10:_j3}
can be seen as using LSB steganography plus some kind of histogram
compensation to remediate the lack of histogram preservation in the
baseline technique.  These beginnings limit their possibilities. For
instance, all of the methods just cited possess a low embedding
rate. The report in~\cite{provos01:_defending} suggests that the
average embedding rate of Outguess lies around 0.31 bits/host element,
whereas Franz's method is below 0.20 bits/host element for most of the
hosts tested in~\cite{franz03:_stegano}, and J3 offers rates between
0.35 bits/host element and 0.65 bits/nonzero host
element~\cite{kumar10:_j3}, all clearly below the ceiling embedding
rate of 1 bit/host element implemented by LSB steganography.

The exception among all histogram-preserving works in terms of
generality and embedding rate is the already mentioned proposal by
Ryabko and Ryabko~\cite{ryabko09:_asymp_optim}, which realizes the
following fundamental observation: any information-carrying
vector~$\mathbf{y}$ that preserves the histogram of~$\mathbf{x}$ must
be a rearrangement of~$\mathbf{x}$. This is because histogram
preservation implies that $\sum_{i=1}^n \ind{v_k=y_i}=\sum_{i=1}^n
\ind{v_k=x_i}$ for all $k=1,2,\ldots,q$, which can only be true if
$\mathbf{y}=\Pi_{\boldsymbol\sigma}\,\mathbf{x}$ for some
permutation~$\boldsymbol\sigma\in\mathcal{S}_n$. This observation
means that the perfect steganographic codes for finite memoryless
hosts are the Variant~I permutation codes first described by
Slepian~\cite{slepian65:permutation}.  This observation was first made
by Franz~\cite{franz03:_stegano}, but Ryabko and
Ryabko~\cite{ryabko09:_asymp_optim} pursued it further in their
proposal of asymptotically optimum perfect steganography. However
their method relies on the algorithm in~\cite{ryabko98:_fast_enumer}
for enumerating combinatorial objects with $O((\log n)^c)$ time
complexity, which, although implementable in special cases, has exponential
memory requirements in general.

As for other algorithms which are explicitly based on rearrangements,
their embedding rate is limited by the fact that they do not exploit
the whole spectrum of histogram-preserving possibilities. For
instance~\cite{luo11:_zero_kullb_liebl}, based on permuting pairs of
host elements, can only achieve a maximum rate of 0.5 bits/host
element. We must also mention that
Mittelholzer~\cite{mittelholzer99:info_theor} was the first to
consider Slepian's permutation modulation as a steganographic
tool. However he %
studied the non-histogram-preserving case
$\mathbf{y}=\mathbf{x}+\Pi_{\boldsymbol\sigma}\mathbf{k}$, with
$\mathbf{k}$ a secret vector, which is not relevant to our problem.

Finally, we outline how previous histogram-preserving approaches have
dealt with the embedding distortion issue. Ryabko and
Ryabko~\cite{ryabko09:_asymp_optim} did not consider any such
constraints. Other histogram-preserving
methods~\cite{provos01:_defending,franz03:_stegano,kumar10:_j3,luo11:_zero_kullb_liebl}
do implement embedding distortion control but only in ad hoc ways, and
so they are not generally amenable to rate-distortion trade-off
optimization.

\section{Permutation Codes as Steganographic Codes}
\label{sec:permutation-codes}
We firstly explore the implications of taking to its full extent the
previous observation about optimum histogram-preserving steganography
of finite memoryless sources necessarily involving Slepian's
permutation coding with $\mathbf{x}$ as the base codeword. Observe
that the central difference with respect to the use of permutation
codes in channel/source
coding~\cite{slepian65:permutation,berger72:_permut} is
that~$\mathbf{x}$ is not a design choice here, but a fixed input
parameter of the encoder (see
Figure~\ref{fig:general}). %
As we will see, this fact is at the root of the two most relevant
challenges we will deal with: encoder complexity and embedding
distortion control.

\subsection{Embedding rate} 
\label{sec:embedding-rate-deg-eff}
We state next some basic facts and definitions about permutation coding
which we will use throughout the paper. The number $r$ of
rearrangements of~$\mathbf{x}$ only depends on its
histogram~$\mathbf{h}$, and it is given by the following multinomial
coefficient:
\begin{equation}
  \label{eq:multi}
  r\triangleq {n\choose \mathbf{h}}=\frac{n!}{h_1! h_2!\cdots h_q!}.
\end{equation}
Hence the $r$ rearrangements
$\mathbf{y}^{(1)},\mathbf{y}^{(2)},\ldots,\mathbf{y}^{(r)}$ of the
host $\mathbf{x}$ are the only codewords that the encoder can produce.
Hereafter we will consider
$\mathcal{S}_\mathbf{x}\subset\mathcal{S}_n$ to be any set of
permutations leading to the $r=|\mathcal{S}_\mathbf{x}|$
rearrangements of~$\mathbf{x}$.

The embedding rate $\rho= (1/n)\log r$ bits/host element associated to
the permutation code based on host $\mathbf{x}$ is necessarily optimum
for first-order perfect steganography, as its calculation incorporates
all possible codewords. In order to obtain a probabilistic perspective
of this quantity, assume next that Stirling's approximation
$\log_e z!\approx z\log_e z- z$ (for large $z$) holds for all
factorials in~\eqref{eq:multi}. In this case the rate can be
informally approximated as
\begin{equation*}
  \rho\approx - \sum_{k=1}^q \frac{h_k}{n} \log \frac{h_k}{n}\quad
  \text{bits/host element}.  \label{eq:rate_entr}
\end{equation*}
If $X$ is a discrete random variable whose probability mass function
is the type of $\mathbf{x}$, $X\sim \mathbf{p}\triangleq
(1/n)\mathbf{h}$, then
\begin{equation}
  \rho\approx \H(X). \label{eq:rhoH}
\end{equation}
This interpretation of the multinomial coefficient in terms of the
entropy of the subjacent type has been known since the definition of
entropy; in the context of permutation coding it was first mentioned
by Berger et al.~\cite{berger72:_permut}. More rigorously, approximation~\eqref{eq:rhoH} is supported by the
following bounds (see for instance~\cite{cover06:elements}):
\begin{equation}
  \H(X)-\zeta^{(q,n)}\le \rho\le \H(X),\label{eq:up_low_bounds_rho}
\end{equation}
where $\zeta^{(q,n)}\triangleq (q/n)\log(n+1)$, which show that
$\rho\to \H(X)$ as $n\to\infty$, i.e., $\rho\doteq \H(X)$. The upper
bound in~\eqref{eq:up_low_bounds_rho} on the capacity of perfect
steganography was previously given by Cachin~\cite{cachin98:inform}
using the method of types, and by Comesaña and
Pérez-González~\cite{comesana07} and Wang and
Moulin~\cite{wang08:_perfec_secur} departing from Gel'fand and
Pinskers' capacity formula, whereas Ryabko and
Ryabko~\cite{ryabko09:_asymp_optim} gave a lower bound alternative to
the one in~\eqref{eq:up_low_bounds_rho}. %

\subsection{Asymptotically optimum encoding  algorithm}
\label{sec:near-optim-embedd}
Even for moderate $n$, the exponentially growing number of
rearrangements of $\mathbf{x}$ precludes the implementation of a naive
encoding scheme, such as a look-up table mapping messages to
rearrangements. Therefore an efficient method to encode messages into
rearrangements (unranking) and to decode messages from rearrangements
(ranking) is essential for a practical implementation of permutation
coding in steganography.

\begin{figure*}[t!]
  \centering
  \begin{subfigure}{0.5\textwidth}
  \centering
  \includegraphics[height=2.8cm]{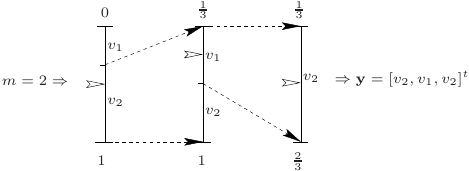}
  \caption{Permutation encoding example}
\end{subfigure}~~~~\begin{subfigure}{0.5\textwidth}
  \centering
  \includegraphics[height=2.8cm]{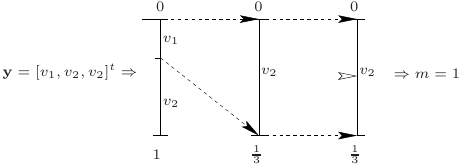}
  \caption{Permutation decoding example}
\end{subfigure}
\caption{Toy example illustrating the use of adaptive arithmetic
  coding to undertake permutation encoding and decoding of vectors
  with histogram $\mathbf{h}=[1,2]^t$ and support
  $\mathbf{v}=[v_1,v_2]^t$, for which $r=3$. The three stages depicted
  represent the right-open intervals $\mathcal{I}^{(0)},
  \mathcal{I}^{(1)}$, and $\mathcal{I}^{(2)}$. The white arrow gives
  the position $(m-1/2)/r$ of message $m$ in $[0,1)$.}
\label{fig:toy}
\end{figure*}

We will describe next a general encoding procedure which is
implementable with $O(n)$ complexity, and which explicitly uses the
duality between optimum perfect universal steganography and optimum
lossless universal source coding. Consider the lossless compression of
a realization~$\mathbf{x}$ of a memoryless signal whose
statistics~$\mathbf{p}$ (or histogram $\mathbf{h}$) are known
beforehand. It is well known that, using the static model~$\mathbf{h}$
of the counts of the support symbols~$\mathbf{v}$ in~$\mathbf{x}$,
arithmetic coding~\cite{pasco76:source,rissanen76:_gener_kraft} can
compress $\mathbf{x}$ with $O(n)$ complexity to the asymptotic
compression rate $H(X)$ (as $n\to\infty$). In a finite context, the
compression rate can be improved by adaptively updating the model
after each symbol in $\mathbf{x}$ is processed by the arithmetic
encoder~\cite{cleary84:comp_enum_adap}, so as to reflect the updated
distribution of the symbols yet to be encoded. In this case the model
is initially $\mathbf{h}$, as in the static case, but after a symbol
is encoded its corresponding count is decremented by one. This
adaptive procedure leads to the implementation of enumerative encoding
by means of arithmetic coding. Every sequence $\mathbf{y}$ with the
same histogram $\mathbf{h}$ as $\mathbf{x}$, i.e.,
$\mathbf{y}=\Pi_{\boldsymbol\sigma}\mathbf{x}$, is compressed to a
sequence that can be seen as the index enumerating $\mathbf{y}$ in the
lexicographic ordering established
by~$\mathbf{v}$. %
The theoretical equivalence between enumerative coding and arithmetic
coding was originally shown by Rissanen~\cite{rissanen79:_arith}, and
Cleary and Witten demonstrated this equivalence in a constructive way by
putting forward the decrementing adaptive model
idea sketched above~\cite{cleary84:comp_enum_adap}.

For the sake of clarity, we will explicitly describe the enumerative
encoding scheme before discussing its role in optimum perfect
universal steganography. Assume that we wish to
compress~$\mathbf{y}=\Pi_{\boldsymbol\sigma}\mathbf{x}$ using
arithmetic coding with the adaptive model discussed above. The initial
right-open interval for the arithmetic encoder is
$\mathcal{I}^{(0)}\leftarrow [0,1)$, and we also make the initialization
$\mathbf{h}^{(0)}\leftarrow\mathbf{h}$.~Then the $i$-th arithmetic
encoding stage (for $i=1,2,\ldots,n$) comprises the following three
steps:
\begin{enumerate}
\item Division step: $\mathcal{I}^{(i-1)}$ is exactly divided into
  nonoverlapping right-open subintervals whose lengths are the nonzero
  fractions $\mathbf{h}^{(i-1)}/(n-i+1)$ of the length of
  $\mathcal{I}^{(i-1)}$. Each subinterval for which $h_k^{(i-1)}>0$ is
  labeled with symbol $v_k$ from $\mathbf{v}$.
\item Encoding step: the subinterval whose label $v_s$ is equal to
  $y_i$ is declared to be the next interval $\mathcal{I}^{(i)}$.
\item Adaptation step: let $h^{(i-1)}_s\leftarrow h^{(i-1)}_s-1$ and
  then declare $\mathbf{h}^{(i)}\leftarrow\mathbf{h}^{(i-1)}$.
\end{enumerate}
If we denote the length of $\mathcal{I}^{(i)}$ by
$|\mathcal{I}^{(i)}|$ then, for any
$\mathbf{y}=\Pi_{\boldsymbol\sigma}\mathbf{x}$, it always holds for
the final interval that $|\mathcal{I}^{(n)}|=\prod_{i=0}^{n-1}
|\mathcal{I}^{(i)}|=1/{n \choose \mathbf{h}}=1/r$. Also, by
construction, the final intervals are always nonoverlapping for any
two different rearrangements $\mathbf{y}$ of $\mathbf{x}$. Thus the
$\lceil\log r\rceil +1$ most significant fractional bits of the binary
representation of the midpoint of~$\mathcal{I}^{(n)}$ constitute the compressed
representation of~$\mathbf{y}$ (Shannon-Fano-Elias coding), or its
index from the viewpoint of enumerative encoding. This representation
can also be put as the most significant $\lceil\log r\rceil+1$
fractional bits of the binary representation of $(m-1/2)/r\in[0,1)$, for some
$m\in\{1,2,\ldots,r\}$. Decompressing $\mathbf{y}$ from~$m$ requires
the same initialization and the same division and adaptation steps as
above, whereas the $i$-th decoding step involves declaring the decoded
symbol~$y_i$ to be equal to the symbol~$v_s$ which labels the
subinterval where $(m-1/2)/r$ lies.  %

We are now ready to outline the permutation coding algorithm, which is
simply the dual of the enumerative encoding algorithm just
explained. The permutation encoder obtains the
rearrangement~$\mathbf{y}=e(\mathbf{x},m)$ by carrying out adaptive
arithmetic decoding of $m$ as described above. On the other hand, the
permutation decoder retrieves the message $m$ embedded
in~$\mathbf{y}$, that is, $m=d(\mathbf{y})$, by carrying out adaptive
arithmetic encoding of~$\mathbf{y}$. The decrementing adaptive model
guarantees that $\mathbf{y}=\Pi_{\boldsymbol\sigma}\mathbf{x}$ for
some $\boldsymbol\sigma\in\mathcal{S}_\mathbf{x}$. Crucially, the
permutation encoder and the permutation decoder share $\mathbf{h}$
---the essential piece of information required for encoding and
decoding--- precisely because of this fact. This is an important
difference with respect to the use of enumerative encoding in
compression, where the encoder must send the model $\mathbf{h}$ along
with the index to the decoder. The permutation encoding and decoding
procedures are illustrated in Figure~\ref{fig:toy}.

Some words are in order about the implementability of the
algorithm. Since $\lceil\log r\rceil+1 > \log r$, the permutation
encoder can map some different messages to the same rearrangement,
which leads to unsolvable ambiguities at the decoder. Univocal
decoding is only guaranteed if at most $\lfloor\log r\rfloor$ bits are
used to represent the messages. This is not a serious limitation
because~\eqref{eq:multi} is usually very large in steganographic
applications, and then $2^{\lfloor \log r\rfloor}\approx r$.

Finally, we would like to remark that the algorithm is closely related
to the one proposed by Berger et al.~\cite[Appendix
VI]{berger72:_permut} in the context of the application of permutation
codes to source coding with a distortion
constraint. %
Although the complexity of the algorithm in~\cite{berger72:_permut} is
claimed to be $O(n)$, it is based on Jelinek's implementation of
Shannon-Fano-Elias coding, and therefore the claim can
only be true for small~$n$ as indicated by Pasco~\cite[page
11]{pasco76:source}. %

\subsubsection*{Keyed encoding}
\label{sec:keyed-encoding}
A way for incorporating a symmetric secret key
$k\in\mathcal{K}\subset\mathbb{N}$ into the algorithm above is to
use~$k$ to select a permutation $\boldsymbol\varsigma\in\mathcal{S}_q$
to be applied to the vectors $\mathbf{h}$ and~$\mathbf{v}$, which are
implicitly shared by encoder and decoder. In other words, the
algorithm stays the same, but encoder and decoder use
$\mathbf{h}^{{\boldsymbol\varsigma}}=\Pi_{\boldsymbol\varsigma}\mathbf{h}$
and
$\mathbf{v}^{{\boldsymbol\varsigma}}=\Pi_{\boldsymbol\varsigma}\mathbf{v}$
instead of $\mathbf{h}$ and~$\mathbf{v}$. With this strategy
$|\mathcal{K}|=q!$ because all values in~$\mathbf{v}$ are unique, even
though this is not true in general for~$\mathbf{h}$. If we represent
the key using $\lfloor\log q!\rfloor$ bits, then
$\mathbf{h}^{\boldsymbol\varsigma}$ and
$\mathbf{v}^{\boldsymbol\varsigma}$ can also be found by means of the
permutation coding algorithm that we have described. Finally,
$|\mathcal{K}|$ can be increased by choosing~$t$ permutations
$\boldsymbol\varsigma^{(1)},\boldsymbol\varsigma^{(2)},\ldots,\boldsymbol\varsigma^{(t)}\in\mathcal{S}_q$
and then using $\boldsymbol\varsigma^{(((i-1)\!\!\mod t)+1)}$ in the
$i$-th arithmetic coding stage.

\section{Embedding Distortion}
\label{sec:embedd-dist-geom}

In this section we will analyze the theoretical embedding distortion
induced by permutation coding, and its connections to the embedding
rate.  For the time~being we will not occupy ourselves with the
practical question of how to control the embedding
distortion. However, as we will see in
Section~\ref{sec:embedd-dist-contr}, the analysis that follows will be key for a
practical implementation of distortion-constrained permutation coding.

\subsection{Squared Euclidean distance}\label{sec:squar-eucl-dist}
A useful way to measure the embedding distortion is by means of the
squared Euclidean distance between a codeword~$\mathbf{y}$ and the
host $\mathbf{x}$.  In this case the similarity measure in
Figure~\ref{fig:general} is
$s(\mathbf{x},\mathbf{y})=\|\mathbf{y}-\mathbf{x}\|^2=\|\mathbf{w}\|^2$,
which is the squared 2-norm of the watermark. We will also refer to it
as the power of~$\mathbf{w}$. The main reasons for considering this
embedding distortion measure are the following ones: 1) it enables
direct comparisons with prior research results when normalized by the
power of the host, which yields communications-like signal to noise
ratios widely adopted in data hiding (see
Section~\ref{sec:norm-embedd-dist}); 2) it is amenable to analysis
and, as it will be seen, it provides relevant insights about the use
of permutation coding in steganography, both in terms of geometry and
of rate-distortion properties; and 3) if $\mathbf{x}$ is the product
of a decorrelating unitary linear transform, then $\|\mathbf{w}\|^2$
is preserved in the original (correlated) domain.

Using the fact that all histogram-preserving codewords~$\mathbf{y}$
have the same 2-norm $\|\mathbf{y}\|=\|\mathbf{x}\|$, the power of a
histogram-preserving watermark can be put as
\begin{equation}\label{eq:wm}
  \|\mathbf{w}\|^2=2\,(\|\mathbf{x}\|^2-\mathbf{x}^t\mathbf{y})=2\,(\|\mathbf{x}\|^2-\mathbf{x}^t\Pi_{\boldsymbol\sigma}\mathbf{x})
\end{equation}
for some $\boldsymbol\sigma\in\mathcal{S}_\mathbf{x}$. Since this
distortion is dependent on the message associated to
$\boldsymbol\sigma$, we will derive next two relevant
message-independent embedding distortion measures, which will be seen
to completely suffice in order to approximate and/or
bound~\eqref{eq:wm} for any~$\boldsymbol\sigma$.
\begin{itemize}
\item \textit{Average watermark power.} If the encoder chooses
  messages uniformly at random, then the average watermark power is
  $\overline{\|\mathbf{w}\|^2}\triangleq (1/r)\sum_{m=1}^{r}\|\mathbf{w}^{(m)}\|^2=2\,\left(\|\mathbf{x}\|^2-(1/r)\sum_{\boldsymbol\sigma\in\mathcal{S}_\mathbf{x}}\mathbf{x}^t\
    \Pi_{\boldsymbol\sigma}\mathbf{x}\right)$.  Using next
  expressions~\eqref{eq:sum_per} and~\eqref{eq:epix} from the
  Appendix, and observing that
  $\mathbf{x}^t\mathbf{1}\mathbf{1}^t\mathbf{x}=(\mathbf{x}^t\mathbf{1})^2$,
  one arrives at
\begin{equation}\label{eq:wm_av_2}
  \overline{\|\mathbf{w}\|^2}=2\,\left(\|\mathbf{x}\|^2-\frac{1}{n}(\mathbf{x}^t\mathbf{1})^2\right).
\end{equation}
Since $(\mathbf{x}^t\mathbf{1})^2\ge 0$ it holds that
\begin{equation}
  \overline{\|\mathbf{w}\|^2}\le 2 \|\mathbf{x}\|^2,\label{eq:lb_av_wm}
\end{equation}
with equality for any zero-sum~$\mathbf{x}$. As we will see, the average watermark power plays a pivotal role in the
application of permutation coding to steganography. For a start, we
verify next that~\eqref{eq:wm_av_2} indicates already that permutation
coding satisfies fundamental theoretical requirements of perfect
steganography. The average watermark power per host element can be put
as
\begin{equation}
  \label{eq:emp_var}
   \frac{1}{n}\overline{\|\mathbf{w}\|^2}=2\;\sigma^2_\mathbf{x},
\end{equation}
where $\sigma^2_\mathbf{x}$ is the (biased) sample variance
of~$\mathbf{x}$. Since the maximum embedding rate is achieved when the
encoder is free to generate all rearrangements of $\mathbf{x}$,
then~\eqref{eq:emp_var} is the exact coding analogous of the
theoretical result by Comesaña and Pérez-González in~\cite[page
17]{comesana07} showing that the average quadratic embedding
distortion in unconstrained capacity-achieving perfect
steganography is
\begin{equation}
  \frac{1}{n}\E\{\|\boldsymbol{W}\|^2\}=2\Var\{X\},\label{eq:pedrofernando}
\end{equation}
where $X$ is a random variable describing an independent and
identically distributed (i.i.d.) host, and $\boldsymbol{W}$ is a
random $n$-vector describing a perfect watermark.

\item \textit{Maximum watermark power.} It is also desirable to obtain
  the maximum power of a perfect steganographic watermark,
  $\|\mathbf{w}\|^2_\mathrm{max}\triangleq \max_{m\in\{1,2,\ldots,r\}}
  \|\mathbf{w}^{(m)}\|^2$, which is the worst-case embedding
  distortion. To this end we may use the following rearrangement
  inequality~\cite[Chapter~10]{hardy34:inequalities}:
\begin{equation}
  {{\mathbf{r}^\uparrow}}^t{\mathbf{s}^\downarrow}\le
  \mathbf{r}^t\mathbf{s}, \label{eq:hardy}
\end{equation}
which holds for any $\mathbf{r},\mathbf{s}\in\mathbb{R}^n$. Setting
$\mathbf{r}=\mathbf{x}$ and $\mathbf{s}=\mathbf{y}$, as~$\mathbf{y}=\Pi_{\boldsymbol\sigma}\mathbf{x}$ we have
from~\eqref{eq:wm} and~\eqref{eq:hardy} that
\begin{eqnarray}\label{eq:wmax}
  \|\mathbf{w}\|^2_\mathrm{max}=2\left({\|\mathbf{x}\|^2-{\mathbf{x}^\uparrow}}^t{\mathbf{x}^\downarrow}\right).
\end{eqnarray}
An upper bound on $\|\mathbf{w}\|^2_\mathrm{max}$ is not immediately
apparent from inspecting~\eqref{eq:wmax}, because
${\mathbf{x}^\uparrow}^t{\mathbf{x}^\downarrow}$ may be
negative. Since $\mathrm{J}=\mathrm{J}^t$, in order to bound
$\|\mathbf{w}\|^2_\mathrm{max}$ from above we can use the
Rayleigh-Ritz theorem to write
${\mathbf{x}^\uparrow}^t{\mathbf{x}^\downarrow}={\mathbf{x}^\uparrow}^t\mathrm{J}{\mathbf{x}^\uparrow}\ge
\lambda_\mathrm{min}(\mathrm{J}) \|\mathbf{x}\|^2$, where
$\lambda_\mathrm{min}(\mathrm{J})$ is the minimum eigenvalue of
$\mathrm{J}$. By definition, an eigenvalue $\lambda$ of $\mathrm{J}$
and its eigenvector~$\mathbf{u}$ fulfill
$\mathrm{J}\mathbf{u}=\lambda\mathbf{u}$. Multiplying this expression
by $\mathbf{u}^t$ one obtains
$\mathbf{u}^t\mathrm{J}\mathbf{u}=\lambda\|\mathbf{u}\|^2$;
alternatively, multiplying it by $\mathbf{u}^t\mathrm{J}$ one obtains
$\|\mathbf{u}\|^2=\lambda\mathbf{u}^t\mathrm{J}\mathbf{u}$, because
 $\mathrm{J}$ is involutory
($\mathrm{J}\mathrm{J}=\mathrm{I}$). Combining these two equations one
sees that $\lambda^2=1$, and since $\tr \mathrm{J}\in\{0,1\}$ then
$\lambda_\mathrm{min}(\mathrm{J})=-1$ when $n>1$. Therefore
\begin{equation}
  \|\mathbf{w}\|^2_\mathrm{max}\le 4 \|\mathbf{x}\|^2.\label{eq:ub_max_wm}
\end{equation}
As we will see in Section~\ref{sec:geometry}, this inequality can be
more directly obtained through geometric arguments; we will discuss when
equality occurs in~\eqref{eq:ub_max_wm} in that section.

\end{itemize}
Finally, a basic inequality involving~\eqref{eq:wm_av_2}
and~\eqref{eq:wmax} is
\begin{equation}
  \overline{\|\mathbf{w}\|^2} \le \|\mathbf{w}\|^2_\mathrm{max}, \label{eq:avmax}
\end{equation}
in which equality occurs when $\mathbf{x}=v\mathbf{1}$. In this case
$r=1$, and so $\overline{\|\mathbf{w}\|^2} = \|\mathbf{w}\|^2_\mathrm{max}=0$.

\subsubsection{Power ratios}
\label{sec:norm-embedd-dist}
The embedding distortion expressions \eqref{eq:wm_av_2}
and~\eqref{eq:wmax} must be normalized in order to be meaningful
across different hosts. Thus, the following figures of merit for the
theoretical embedding distortion can be put forward:
\begin{itemize}
\item Host to average watermark power ratio
  \begin{equation}
    \underline{\xi}\triangleq \frac{\|\mathbf{x}\|^2}{\overline{\|\mathbf{w}\|^2}}.
    \label{eq:pdawr_def}
  \end{equation}
  A common alternative to $\underline{\xi}$ is the peak host to
  average watermark power ratio $\underline{\xi}'\triangleq
  n(2^b-1)^2/\overline{\|\mathbf{w}\|^2}$, where it is assumed that
  $\mathbf{x}$ is in the nonnegative orthant and represented
  using~$b$~bits/element.
\item Host to maximum watermark power ratio
  \begin{equation*}
    \xi_\mathrm{min}\triangleq   \frac{\|\mathbf{x}\|^2}{\|\mathbf{w}\|^2_\mathrm{max}}.\label{eq:dwcwr_def}
\end{equation*}
\end{itemize}
These figures of merit are  related as follows:
\begin{equation}
  \underline{\xi}'\ge \underline{\xi}\ge \xi_\mathrm{min}. \label{eq:ineq_fig_merit}
\end{equation}
In keeping with standard conventions and where convenient throughout
the paper, we will also refer to $\xi$ ratios in terms of decibels
(dB), by which the amount $10\log_{10}\xi$ is understood. For all the
convenience of these figures of merit, a word of caution is needed: as
a rule, low $\xi$ ratios imply dissimilarity between $\mathbf{x}$ and
$\mathbf{y}$ (or of their counterparts in the original domain if
decorrelation is used to approximate the memoryless hypothesis), but
the converse is not always true.  An example of this shortcoming is
the widely used peak signal to noise ratio (PSNR) of which
$\underline{\xi}'$ is a version.  According to our discussion, high
$\xi$ ratios are a necessary condition to ensure
similarity. However~\eqref{eq:lb_av_wm} implies
that~$\underline{\xi}\ge 1/2$ ($\approx -3$~dB),
whereas~\eqref{eq:ub_max_wm} implies that~$\xi_\mathrm{min}\ge 1/4$
($\approx -6$~dB). The fact that both minima are very low makes it
clear that a mechanism for embedding distortion control will be
required whenever similarity must be enforced, which will be dealt
with in Section~\ref{sec:embedd-dist-contr}.

Finally, our discussion about the analogy between~\eqref{eq:emp_var}
and~\eqref{eq:pedrofernando} also let us see why the host to average
embedding distortion ratio of capacity-achieving perfect steganography
for zero-mean~$X$ in~\cite{comesana07} coincides
with~$\underline{\xi}$ for zero-sum~$\mathbf{x}$, that is to say,
$\underline{\xi}\approx -3$~dB.

\subsubsection{Asymptotics}
\label{sec:asymptotic}
We will study next the asymptotic behavior for large $n$ of the power
a histogram-preserving watermark drawn uniformly at random,
corresponding to the encoder choosing messages uniformly at
random. Our aim is to quantify a condition under
which~\eqref{eq:wm_av_2} is a good predictor of the power of
\textit{any} histogram-preserving watermark, via the weak law of large
numbers.  In order to do so we will obtain Chebyshev's inequality for
the random variable
$\|\boldsymbol{W}\|^2=2(\|\mathbf{x}\|^2-\mathbf{x}^t\mathit{\Pi}\mathbf{x}$),
defined using~\eqref{eq:wm} and assuming that $\mathit{\Pi}$ is a
random variable uniformly distributed over a set of permutation
matrices of cardinality $|\mathcal{S}_\mathbf{x}|$ that can generate all
rearrangements of~$\mathbf{x}$. We know already that
$\E\{\|\boldsymbol{W}\|^2\}=\overline{\|\mathbf{w}\|^2}$, and
therefore we just need to obtain the second moment
of~$\|\boldsymbol{W}\|^2$ to compute its variance. As
$\mathbf{x}^t\Pi\mathbf{x}=\mathbf{x}^t\Pi^t\mathbf{x}$, this moment
can be put as
\begin{eqnarray*}
  \label{eq:ew4}
  \E\!\left\{\|\boldsymbol{W}\|^4\right\}
  &=&4\,\mathbf{x}^t\left(\mathbf{x}\mathbf{x}^t-2\|\mathbf{x}\|^2\E\{\mathit{\Pi}\}+\E\!\left\{\mathit{\Pi}\mathbf{x}\mathbf{x}^t\mathit{\Pi}^t\right\}\right)\mathbf{x},
\end{eqnarray*}
and hence the desired variance is
\begin{eqnarray}
  \label{eq:varw2}
  \hspace*{-.5cm}\Var\left\{\|\boldsymbol{W}\|^2\right\}
  &=&4\left(\mathbf{x}^t\E\!\left\{\mathit{\Pi}\mathbf{x}\mathbf{x}^t\mathit{\Pi}^t\right\}\mathbf{x}-\left(\mathbf{x}^t\E\{\mathit{\Pi}\}\mathbf{x}\right)^2\right).
\end{eqnarray}
Using the computation of the two expectations in~\eqref{eq:varw2}
found in the Appendix, it can be seen after some algebraic
manipulations that~\eqref{eq:varw2} becomes
\begin{eqnarray*}
  \label{eq:varw22}
  \Var\left\{\|\boldsymbol{W}\|^2\right\}
  &=&\frac{1}{n-1}\left(\overline{\|\mathbf{w}\|^2}\right)^2. %
\end{eqnarray*}
Finally, we use $\Var\{\|\boldsymbol{W}\|^2\}$ and
$\E\{\|\boldsymbol{W}\|^2\}$ in Chebyshev's inequality for
$\|\boldsymbol{W}\|^2$. Given $\gamma>0$, this inequality can be put as
$\Pr\{|Z-\mu|\ge \gamma \mu\}\le \sigma^2/(\gamma^2\mu^2)$ for a
variable~$Z$ with mean~$\mu$ and variance~$\sigma^2$. Thus we obtain
\begin{equation}
  \label{eq:chebyshev}
  \Pr\left\{\left|\|\boldsymbol{W}\|^2-\overline{\|\mathbf{w}\|^2}\right|\ge\gamma\overline{\|\mathbf{w}\|^2}\right\}\le  \frac{1}{\gamma^2(n-1)}.
\end{equation}
Although Chebyshev's inequality is
known to be loose
it
is also completely general, and it can be read as saying that the
embedding distortion associated to a randomly drawn permutation
codeword is not likely to be too different from the
average~$\overline{\|\mathbf{w}\|^2}$ for large~$n$. This fact will be
empirically verified in Section~\ref{sec:results}.

Despite these considerations, one might still be concerned about the
rare instances in which $\|\boldsymbol{W}\|^2$  deviates from
the average. We will see next that the geometry of permutation coding
strictly confines the maximum distortion~\eqref{eq:wmax} in terms
of the average distortion~\eqref{eq:wm_av_2}.

\subsubsection{Geometry}
\label{sec:geometry}
As noted by Slepian~\cite{slepian65:permutation}, the two basic
observations to be made about the geometry of permutation codes are:
1) since $\|\mathbf{y}\|=\|\mathbf{x}\|$, then all codewords lie on an
$n$-dimensional primary permutation sphere centered at~$\mathbf{0}$
with radius $\|\mathbf{x}\|$; and 2) the codewords are really $n-1$ dimensional,
as they also lie on the permutation plane with equation
$\mathbf{y}^t\mathbf{1}=\mathbf{x}^t\mathbf{1}$.

As we will show next, relevant geometric insights for the embedding
distortion analysis can be obtained from what we will call the
secondary permutation sphere\footnote{It is possible to prove that the
  secondary permutation sphere is also the covering sphere of the
  permutation code. For the sake of brevity we omit the proof, as it
  is not consequential for our analysis.}. This is a sphere with
center $\mathbf{c}$ in the permutation plane
($\mathbf{c}^t\mathbf{1}=\mathbf{x}^t\mathbf{1}$) and radius~$R_s$
such that $\|\mathbf{y}-\mathbf{c}\|^2=R_s^2$ for any
codeword~$\mathbf{y}$. Since the intersection of the primary permutation
sphere with the permutation plane is a sphere in $n-1$ dimensions that
contains all codewords, then this locus must coincide with the
intersection of the secondary permutation sphere with the permutation
plane. In order to obtain~$\mathbf{c}$ and~$R_s$ we start by computing
the average of all codewords
$\overline{\mathbf{y}}\triangleq (1/r)\sum_{m=1}^r
\mathbf{y}^{(m)}=(1/r)\sum_{\boldsymbol\sigma\in\mathcal{S}_\mathbf{x}}
\Pi_{\boldsymbol\sigma}\mathbf{x}$.
Using~\eqref{eq:sum_per} and~\eqref{eq:epix} it can be seen that this
average vector is
\begin{equation*}
  \label{eq:average}
  \overline{\mathbf{y}}=\frac{1}{n}(\mathbf{x}^t\mathbf{1})\mathbf{1}.
\end{equation*}
So all coordinates of $\overline{\mathbf{y}}$ equal the average
of~$\mathbf{x}$. As indicated by
$\overline{\mathbf{y}}^t\mathbf{1}=\mathbf{x}^t\mathbf{1}$,
$\overline{\mathbf{y}}$ lies on the permutation plane.  Now, the
square of the Euclidean distance of an arbitrary codeword $\mathbf{y}$ to
$\overline{\mathbf{y}}$ is
\begin{equation}
  \|\mathbf{y}-\overline{\mathbf{y}}\|^2=\|\mathbf{x}\|^2-\|\overline{\mathbf{y}}\|^2,\label{eq:dist_y_avy}
\end{equation}
where we have used $\mathbf{y}^t\mathbf{1}=\mathbf{x}^t\mathbf{1}$ and
$\mathbf{y}^t\overline{\mathbf{y}}=(1/n)(\mathbf{x}^t\mathbf{1})^2=\|\overline{\mathbf{y}}\|^2$. As~\eqref{eq:dist_y_avy}
is independent of~$\mathbf{y}$, then it must also be the square of the
secondary permutation sphere radius,~$R_s^2$, and the center of this sphere must
be $\mathbf{c}=\overline{\mathbf{y}}$. From~\eqref{eq:wm_av_2}, we can
thus write
\begin{equation}
  \label{eq:secondary_radius}
  R_s^2=\frac{1}{2}\overline{\|\mathbf{w}\|^2}.
\end{equation}
Therefore when $\mathbf{x}^t\mathbf{1}\neq 0$ all codewords lie
simultaneously on two different spheres: the primary and the secondary permutation spheres. The equation of the plane where these two spheres
intersect must be the permutation plane
$\mathbf{y}^t\mathbf{1}=n
\sqrt{\|\mathbf{x}\|^2-R_s^2}=\mathbf{x}^t\mathbf{1}$.
The secondary permutation sphere radius cannot be greater than the radius of the
primary permutation sphere, as it can be seen, for example,
from~\eqref{eq:secondary_radius} and~\eqref{eq:lb_av_wm}. Therefore
\begin{equation}
  R_s\le \|\mathbf{x}\|\label{eq:rc_nx},
\end{equation}
with equality when $\mathbf{x}^t\mathbf{1}= 0$.  This is the likely reason
why the secondary permutation sphere was never considered in previous works devoted
to the application of permutation codes to channel
coding~\cite{slepian65:permutation} or source
coding~\cite{berger72:_permut}, since in those scenarios
$\mathbf{x}^t\mathbf{1}= 0$ is usually necessary for energy
minimization (and also feasible, since~$\mathbf{x}$ is a chosen parameter in
both problems), and hence the two spheres coincide.
\begin{figure}[t!]
  \centering
  \includegraphics[width=5.8cm]{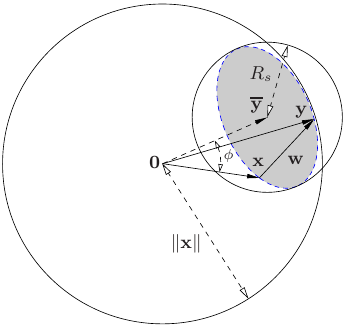}
  \caption{Schematic of the geometry of first-order perfect steganography.}
  \label{fig:spheres}
\end{figure}

Using the triangle inequality we can verify next that
\begin{equation}
  \|\mathbf{w}\|=\|(\mathbf{y}-\overline{\mathbf{y}})-(\mathbf{x}-\overline{\mathbf{y}})\|\le
  2 \|\mathbf{y}-\overline{\mathbf{y}}\|=2
  R_s, \label{eq:triangle1}
\end{equation}
or, equivalently, that $\|\mathbf{w}\|$ cannot be greater than the
diameter of the secondary permutation sphere. Using~\eqref{eq:secondary_radius}, we
see that
\eqref{eq:triangle1} implies the inequality
\begin{equation}
  \|\mathbf{w}\|^2_\mathrm{max}\le 2\overline{\|\mathbf{w}\|^2},
  \label{eq:inequalities}
\end{equation} which supplements~\eqref{eq:avmax}. A case in which
equality holds in~\eqref{eq:inequalities} is~$\mathbf{x}=v\,\mathbf{1}$,
but there may also exist other equality solutions
with a nonconstant
host. %
Inequality~\eqref{eq:inequalities} also narrows down the probability
bound in~\eqref{eq:chebyshev}, since it implies that this
probability can only be nonzero when~$\gamma\in (0,1]$. Finally,
when~\eqref{eq:inequalities} is normalized by the power of the host we
obtain
\begin{equation}
  \label{eq:xi_min_lb}
  \xi_\mathrm{min}\ge \underline{\xi}/2,
\end{equation}
which means that the host to maximum watermark power ratio is, in
decibels, always greater or equal than the host to average watermark
power ratio minus approximately $3$~dB.  %

An alternative proof of inequality~\eqref{eq:ub_max_wm} is found, for
instance, by combining~\eqref{eq:triangle1} and~\eqref{eq:rc_nx}. This
can also be seen directly by applying the triangle inequality as
in~\eqref{eq:triangle1} but with~$\mathbf{0}$ instead
of~$\overline{\mathbf{y}}$, which is equivalent to saying that
$\|\mathbf{w}\|$ cannot be greater than the diameter of the
primary permutation sphere. Thus~\eqref{eq:ub_max_wm} is met with equality
whenever there exist two antipodal codewords, that is to say, when
there exist $\mathbf{y}$ and $\mathbf{y}'$ such that
$\mathbf{y}'=-\mathbf{y}$. This happens when
$\mathbf{x}^\uparrow=-\mathbf{x}^\downarrow$, %
two antipodal codewords being
$\mathbf{y}={\mathbf{x}^\uparrow}$ and
$\mathbf{y}'={\mathbf{x}^\downarrow}$. %
In the special case
in which~$\mathbf{x}$ lies in the nonnegative (or nonpositive)
orthant, the greatest possible diameter of the secondary permutation sphere allows us
to replace~\eqref{eq:ub_max_wm} by $\|\mathbf{w}\|^2_\mathrm{max}\le
2\|\mathbf{x}\|^2$ (and hence $\xi_\mathrm{min}\ge 1/2$), with
equality when at least half of the elements of $\mathbf{x}$ are zero.
  
To conclude this section, it can also be observed that the
host to average watermark power ratio can be expressed as a single
function of the angle~$\phi$ between $\mathbf{x}$ and~$\mathbf{1}$
(equivalently, between any codeword $\mathbf{y}$
and~$\mathbf{1}$). Since $\cos
\phi=\mathbf{x}^t\mathbf{1}/\left(\|\mathbf{x}\|\|\mathbf{1}\|\right)$,
it can be seen from~\eqref{eq:wm_av_2} and~\eqref{eq:pdawr_def} that
\begin{equation}
  \label{eq:dwrq}
  \underline{\xi}=\frac{1}{2\sin^2\phi}.
\end{equation}
This expression also allows us to establish~\eqref{eq:xi_min_lb}
without explicitly resorting to the secondary permutation sphere.  Since the
angle $2\phi$ is the opening angle of the right cone with
apex~$\mathbf{0}$ and base the intersection of the primary permutation sphere
and the permutation plane, then the maximum distance between any
codeword~$\mathbf{y}$ and~$\mathbf{x}$ is bounded as follows:
\begin{equation}
  \|\mathbf{y}-\mathbf{x}\|^2\le 2 \|\mathbf{x}\|^2 (1-\cos 2\phi).\label{eq:alt_inequalities}
\end{equation}
Using next the trigonometric identity $\cos 2\phi=1-2\sin^2 \phi$
and~\eqref{eq:dwrq} in~\eqref{eq:alt_inequalities} we
recover~\eqref{eq:xi_min_lb}.

Some of the facts discussed in this section are schematically
illustrated in Figure~\ref{fig:spheres}.

\subsection{Degree of host change}
\label{sec:average-degree-host}
The degree of host change is an embedding distortion measure
alternative to the squared Euclidean distance, which, as we will see,
is also insightful in a number of ways. In this case the similarity
measure is
$s(\mathbf{x},\mathbf{y})=(1/n)\,\delta(\mathbf{y},\mathbf{x})=(1/n)\omega(\mathbf{w})$,
that is to say, the Hamming distance per symbol between $\mathbf{y}$
and~$\mathbf{x}$, or the fraction of elements of $\mathbf{y}$ which
differ from the same index elements in $\mathbf{x}$. For simplicity, in
the remainder we will use the notation $\nu\triangleq
(1/n)\,\delta(\mathbf{y},\mathbf{x})$.

We will determine next the average degree of host change over all
rearrangements~($\overline{\nu}$), in order to have a measurement
independent of any particular~$\mathbf{y}$. %
We firstly define an auxiliary $q\times n$ matrix $\Lambda$ whose
entries are $(\Lambda)_{k,i}=\ind{v_k=x_i}$, and we let
$\Omega\triangleq\Lambda^t\Lambda$. Now, $\tr
\Lambda\Pi_{\boldsymbol\sigma}\Lambda^t=\tr\Omega\Pi_{\boldsymbol\sigma}$
is the number of elements in
$\mathbf{y}=\Pi_{\boldsymbol\sigma}\mathbf{x}$ unchanged with respect
to~$\mathbf{x}$. Therefore $\delta(\mathbf{y},\mathbf{x})=n-\tr
\Omega\Pi_{\boldsymbol\sigma}$, and, when the messages are uniform,
the average degree of host change can be put as
\begin{equation}
  \overline{\nu}= \frac{1}{r}\sum_{\boldsymbol\sigma\in\mathcal{S}_\mathbf{x}}\frac{1}{n} (n-\tr \Omega\Pi_{\boldsymbol\sigma}). \label{eq:av_degree_host_change}
\end{equation}
We can develop this expression using the equality
$(1/r)\sum_{\boldsymbol{\sigma}\in\mathcal{S}_\mathbf{x}}\tr\Omega\Pi_{\boldsymbol\sigma}=(1/n!)\sum_{\boldsymbol{\sigma}\in\mathcal{S}_n}\tr
\Omega\Pi_{\boldsymbol\sigma}$, which holds because the second
summation contains the same summands as the first one, but each of
them repeated $h_1!h_2!\cdots h_q!$ times.  As the trace operator is
linear, using equation~\eqref{eq:sum_per} and $\tr
\Omega\mathbf{1}\mathbf{1}^t=\mathbf{1}^t\Omega\mathbf{1}=\|\Lambda\mathbf{1}\|^2=\|\mathbf{h}\|^2$,
it can be seen that
\begin{equation}
  \label{eq:avnu}
  \overline{\nu}=1-\frac{\|\mathbf{h}\|^2}{n^2}=1-\|\mathbf{p}\|^2.
\end{equation}
As $\mathbf{p}$ (the type of $\mathbf{x}$) contains the probabilities
of a pmf (whose support is~$\mathbf{v}$) then $1/q\le
\|\mathbf{p}\|^2\le1$, and so we have that
\begin{equation}
  0\le\overline{\nu}\le 1-\frac{1}{q}. \label{eq:nubounds}
\end{equation}
A useful probabilistic interpretation of~\eqref{eq:avnu} can be
obtained as follows. Consider two independent discrete random
variables whose distribution is the type of $\mathbf{x}$, which we
denote as $X\sim \mathbf{p}$ and $Y\sim\mathbf{p}$. The complement of
their \textit{index of coincidence}, or, equivalently, the probability
of drawing a different outcome in two independent trials of $X$
and~$Y$, is $\Pr\{X\neq Y\}=\sum_{v\in\mathcal{V}} p(X=v)p(Y\neq
v)$. This amount is
\begin{eqnarray}\label{eq:icnu}
  \Pr\{X\neq Y\}&=&\sum_{k=1}^q p_k\left(1-p_k\right)=\overline{\nu}.
\end{eqnarray}
As for the relationship between $\overline{\nu}$ and
$\overline{\|\mathbf{w}\|^2}$, since
$\delta(\mathbf{y},\mathbf{x})\le\|\mathbf{y}-\mathbf{x}\|^2$ for
$\mathbf{x},\mathbf{y}\in\mathbb{Z}^n$ then
\begin{equation}\label{eq:nuw}
  \overline{\nu}\le \frac{1}{n} \overline{\|\mathbf{w}\|^2},
\end{equation}
The right-hand side of~\eqref{eq:nuw} can be greater than one, but the
bound is tight when $\underline{\xi}\to\infty$; intuitively, a
high~$\underline{\xi}$ implies a small $\overline{\nu}$.  Finally note
that, unlike the $\xi$ ratios, $\overline{\nu}$ is not preserved, in
general, by unitary linear transforms when $q>2$.

\subsection{Rate-distortion bounds}
\label{sec:rate-dist-bounds}
Ideally we would like to have exact rate-distortion relationships,
that is to say, explicit or implicit equations relating the embedding
rate~$\rho$ discussed in Section~\ref{sec:embedding-rate-deg-eff} and
either $(1/n)\overline{\|\mathbf{w}\|^2}$ or~$\overline{\nu}$. Leaving
aside the asymptotic case of the binary Hamming setting which we
discuss later in Section~\ref{sec:binary-case}, we suspect that, in
general, such relationships do not exist.  If they would, they could
not simply involve the three aforementioned amounts. Two particular
nonbinary hosts validating this observation are $\mathbf{x}=[1, 2, 3,
4, 4, 4, 4]^t$ and $\mathbf{x}'=[1, 1, 2, 2, 3, 3, 3]^t$, which have
the same~$\rho$ but different $(1/n)\overline{\|\mathbf{w}\|^2}$ and
$\overline{\nu}$.

Nevertheless, a general upper bound on $\rho$ solely based on
$(1/n)\overline{\|\mathbf{w}\|^2}$ is possible using the differential
entropy upper bound on discrete entropy, independently found by
Djackov~\cite{djackov77}, Massey~\cite{massey88}, and~Willems
(unpublished, see~\cite[Problem 8.7]{cover06:elements}).  This bound
is $\H(X)< (1/2)\log(2\pi e (\sigma_X^2+1/12))$ for a discrete random
variable $X$ with support set~$\mathcal{V}\subseteq\mathbb{Z}$ and
variance~$\sigma_X^2$. Using the upper bound
in~\eqref{eq:up_low_bounds_rho}, since $X\sim\mathbf{p}$ has support
set $\mathcal{V}\subset\mathbb{Z}$ and variance
$\sigma^2_\mathbf{x}$ [given in~\eqref{eq:emp_var}], we have
that the Djackov-Massey-Willems bound yields
\begin{equation}
  \rho< \rho_u\triangleq \frac{1}{2} \log\left(2\pi e \left(\frac{ \overline{\|\mathbf{w}\|^2}}{2n}+\frac{1}{12}\right)\right).\label{eq:rd-function}
\end{equation}
The bound is loose as $\overline{\|\mathbf{w}\|^2}/n\to 0$, as in this
case $\rho_u\to 0.2546$ whereas we know that $\rho\to 0$. In practical
terms, this effect only starts to be discernible when
$\underline{\xi}'\gtrapprox 50$ dB for $b=8$.  

Also, from~\eqref{eq:icnu} and the upper bound
in~\eqref{eq:up_low_bounds_rho}, another rate-distortion upper bound
is directly given by Fano's inequality:
\begin{equation}
  \label{eq:fano}
  \rho \le \rho_u'\triangleq h(\overline{\nu})+\log(q-1)\,\overline{\nu}.
\end{equation}
Although it is well-known that the sharpest bound is achieved by any
permutation of
$\mathbf{p}=[1-\overline{\nu},\overline{\nu}/(q-1),\cdots,\overline{\nu}/(q-1)]^t$,
note from~\eqref{eq:up_low_bounds_rho} that this does not imply that
$\rho=\rho_u'$. Moreover the type $\mathbf{p}$ is fixed for the
problem, and so this bound is generally loose.

Furthermore, it is possible to give two lower bounds on $\rho$ based
on~$\overline{\nu}$. The first one is based on  inequality $\Pr\{X=Y\}\ge
2^{-\H(X)}$ with $X$ and $Y$ i.i.d.~\cite[Lemma
2.10.1]{cover06:elements}~%
Using the complement of
the index of coincidence~\eqref{eq:icnu} and the lower bound
in~\eqref{eq:up_low_bounds_rho} in this inequality, we obtain
\begin{equation}
  \label{eq:bound-rate-dhc}
  \rho\ge \rho_l\triangleq -\log(1-\overline{\nu})-\zeta^{(q,n)}.
\end{equation}
Asymptotically, we have that $\rho_l\doteq -\log(1-\overline{\nu})$. Since the
original bound is sharp when $X$ is uniform, asymptotic equality is
achieved in inequality~\eqref{eq:bound-rate-dhc} when
$\mathbf{h}=(n/q)\mathbf{1}$, and in this case $\rho=\rho_l\doteq \log q$,
which corresponds to the maximum average degree of host
change~$\overline{\nu}=1-1/q$.

The second lower bound is obtained from inequality $\H(X)\ge 2
\Pr\{X\neq Y\}$ with $X$ and $Y$ i.i.d., found by Harremoës and
Topsøe~\cite[Theorem II.6 with $k=1$]{harremoes01:_inequalities}. Using again the
lower bound in~\eqref{eq:up_low_bounds_rho} and~\eqref{eq:icnu} we
obtain
\begin{equation}
  \label{eq:rd-nu2}
  \rho \ge \rho_l'\triangleq 2\,\overline{\nu}-\zeta^{(q,n)},
\end{equation}
which is sharper than~\eqref{eq:bound-rate-dhc} when
$0<\overline{\nu}<1/2$. The asymptotic bound is now $\rho_l'\doteq
2\overline{\nu}$.

Finally, from~\eqref{eq:rd-function},~\eqref{eq:bound-rate-dhc}
and~\eqref{eq:rd-nu2} we have two upper bounds on $\overline{\nu}$
based on $(1/n)\overline{\|\mathbf{w}\|^2}$ in addition
to~\eqref{eq:nuw}.  If we define $\tau\triangleq
2^{\zeta^{(q,n)}}(2\pi e (\overline{\|\mathbf{w}\|^2}/2n+1/12))^{1/2}
$, then from $\rho_l<\rho_u$ and from $\rho_l'<\rho_u$ we respectively
have that
\begin{equation}
  \label{eq:bound_nu}
  \overline{\nu}< 1-\frac{1}{\tau},\quad\text{and}\quad
  \overline{\nu}< \frac{1}{2}\log\tau.
\end{equation}
The first upper bound in~\eqref{eq:bound_nu} cannot be greater than
the unity, unlike the second upper bound or~\eqref{eq:nuw}. However,
as $\underline{\xi}\to\infty$~\eqref{eq:nuw} is eventually tighter
than both inequalities in~\eqref{eq:bound_nu}, which is  due to the lack
of asymptotic sharpness of~\eqref{eq:rd-function}.

Since $\rho$, $\overline{\|\mathbf{w}\|^2}$ and $\overline{\nu}$ are
completely determined by $\mathbf{x}$, the rate-distortion bounds
presented in this section may look like little more than a curiosity
at this juncture.  However their relevance will become apparent when
we address embedding distortion control for permutation coding in
Section~\ref{sec:embedd-dist-contr}.

\subsection{Embedding efficiency}
\label{sec:embedding-efficiency}
In this section we will study the average embedding efficiency
($\overline{\varepsilon}$)~\cite{westfeld01:f5}. This quantity is
defined as the average number of message bits embedded per host
element change, and, hence, it simultaneously involves embedding rate
and embedding distortion.  The original idea behind
$\overline{\varepsilon}$ was measuring the security of a
steganographic algorithm: given two algorithms with the same~$\rho$,
the one with higher $\overline{\varepsilon}$ should be less
detectable since, on average, it embeds the same amount of information
with less degree of host change. %

Since permutation coding leads to perfect steganography with finite
memoryless hosts, it may seem that there is little point in
contemplating $\overline{\varepsilon}$ here. However, we will see that
the average embedding efficiency allows for an insightful comparison
between permutation coding and model-based
steganography~\cite{sallee03:model}. Moreover,
$\overline{\varepsilon}$ may also find application in realistic
scenarios in which the memoryless assumption is only an
approximation. The computation of $\overline{\varepsilon}$ will again
require the $\Omega$ matrix defined in
Section~\ref{sec:average-degree-host}. Firstly see that the embedding
efficiency for the message encoded by
$\mathbf{y}=\Pi_{\boldsymbol\sigma}\mathbf{x}$ is
$\varepsilon\triangleq\log r/(n\nu)=\log r/(n-\tr
\Omega\Pi_{\boldsymbol\sigma})$ bits/host element change. This amount
is infinite for $\boldsymbol\sigma_0\in\mathcal{S}_\mathbf{x}$ such
that $\Pi_{\boldsymbol\sigma_0}\mathbf{x}=\mathbf{x}$, which without
loss of generality may be assumed to be $\boldsymbol\sigma_0\triangleq
[1,2,\ldots,n]^t$. In order to sidestep this singularity we will
consider that $\varepsilon=0$ when $\mathbf{y}=\mathbf{x}$. Therefore,
when all messages are equally likely the average sought is 
\begin{equation}
  \overline{\varepsilon}=\frac{1}{r}\sum_{\boldsymbol{\sigma}\in\mathcal{S}_\mathbf{x}\backslash\boldsymbol\sigma_0}\frac{\log
    r}{n-\tr \Omega\Pi_{\boldsymbol\sigma}}\quad\text{bits/host element change}.\label{eq:efficiency}
\end{equation}
An exact evaluation of~\eqref{eq:efficiency} requires enumerating the
number of permutations associated to each possible value of $\tr
\Omega\Pi_{\boldsymbol\sigma}$ for
$\boldsymbol\sigma\in\mathcal{S}_\mathbf{x}\backslash\boldsymbol\sigma_0$,
namely $\{0,1,2,\ldots,n-2\}$ ($n-1$ is not a possible value
because an elementary permutation involves swapping two
indices).  %
Obtaining this enumeration is
equivalent to solving a generalization of the classic problem of
\textit{rencontres}%
~\cite{montmort08:_essay_}, which asks for the number of
$\boldsymbol\sigma\in\mathcal{S}_n$ that exhibit a given number of
fixed points with respect to~${\boldsymbol\sigma}_0$. The generalized
problem at hand requires instead finding the number of rearrangements
of~$\mathbf{x}$ that exhibit a given number of fixed points with
respect to~$\mathbf{x}$. %
We are not aware of a solution to this generalized problem, but a useful
lower bound on~$\overline{\varepsilon}$ can be found by observing
that~\eqref{eq:efficiency} involves the harmonic mean of $r-1$
positive values, which is bounded from above by their arithmetic
mean~\cite{hardy34:inequalities}. Then we have that
\begin{equation}
  \overline{\varepsilon}\ge \overline{\varepsilon}_l\triangleq n\rho\left(\frac{r-1}{r}\right)
  \left(\frac{1}{r-1}\sum_{\boldsymbol{\sigma}\in\mathcal{S}_\mathbf{x}\backslash\boldsymbol\sigma_0}\left(n-\tr
      \Omega\Pi_{\boldsymbol\sigma}\right)\right)^{-1}.\label{eq:efficiency_b}
\end{equation}
As the sum over
$\boldsymbol\sigma\in\mathcal{S}_\mathbf{x}\backslash\boldsymbol\sigma_0$
in~\eqref{eq:efficiency_b} is equal to the same sum over
$\boldsymbol\sigma\in\mathcal{S}_\mathbf{x}$ we can 
see using~\eqref{eq:av_degree_host_change} that
\begin{equation}
  \overline{\varepsilon}_l=
\left(\frac{r-1}{r}\right)^2\,   \frac{\rho}{\overline{\nu}}.\label{eq:average_epsilon}
\end{equation}
Recalling the lower bound in~\eqref{eq:up_low_bounds_rho} we can in
turn bound~\eqref{eq:average_epsilon} from below as follows:
$\overline{\varepsilon}_l\ge \overline{\varepsilon}_l'\triangleq
((r-1)/r)^2 (H(X)-\zeta^{(q,n)})/\overline{\nu}$.
In the following we will consider the asymptotics of $\overline{\varepsilon}_l'$
 as $n\to\infty$. As in this case $(r-1)/r\to 1$ we have
that
\begin{equation}
  \overline{\varepsilon}_l'\doteq 
  \frac{-\sum_{k=1}^q p_k \log p_k}{1-\sum_{k=1}^q p_k^2}.
  \label{eq:asympt_lb}
\end{equation}
A basic but loose lower bound on~\eqref{eq:asympt_lb} can be found by
applying the well-known inequality $p-1\ge \log_e p$ to every
logarithm in the numerator of the expression, which yields
$\overline{\varepsilon}_l'\; \shortstack[c]{\vspace*{-.05cm}.\\$>$}\;
1/(\log_e 2)\approx
1.44$, but the sharpest lower bound is obtained by applying
to~\eqref{eq:asympt_lb} the same inequality
from~\cite{harremoes01:_inequalities} used to obtain~\eqref{eq:rd-nu2},
which yields the asymptotic lower bound
\begin{equation}
  \overline{\varepsilon}_l'\;\shortstack[c]{\vspace*{-.05cm}.\\$\ge$}\; 2. \label{eq:ineq_vareps}
\end{equation}
Consequently, a minimum average embedding efficiency of~$2$~bits/host
element change is asymptotically guaranteed when using permutation
coding.

\subsection{Binary host}
\label{sec:binary-case}
We now particularize and expand some of the previous
results %
for the special case~$q=2$. This case may arise because the host is
intrinsically binary or, as we will discuss in
Section~\ref{sec:embedd-dist-contr}, because we are dealing with a
two-valued partition of a nonbinary host (see
Section~\ref{sec:hist-pres-lsb}).  In the binary case the number of
rearrangements~\eqref{eq:multi} is given by the binomial coefficient
$r={n\choose h_2}=n!/\left((n-h_2)!h_2!\right)$,
and~\eqref{eq:asympt_lb} becomes
\begin{equation*}
  \label{eq:approx_eps_binary}
  \overline{\varepsilon}_l'\doteq \frac{-(1-p_2)\log (1-p_2)-p_2\log p_2}{2\,p_2(1-p_2)}.
\end{equation*}
This same expression was previously given by Sallee~\cite[page
11]{sallee03:model} for the average embedding efficiency of
model-based steganography with quantizer step size~2 (by which Sallee means a
two-valued partitioning of a host signal using adjacent pairs of
histogram bins), although we have proved here that it is only an
asymptotic lower bound.  Inequality~\eqref{eq:ineq_vareps} was also
mentioned in~\cite{sallee03:model}, but the justification therein
is, apparently, only empirical and restricted to the binary case
---note that our conclusion is based on the aforementioned theoretical
result by Harremoës and Topsøe~\cite{harremoes01:_inequalities}, which
holds for arbitrary~$q$. We will return to the connection between
permutation coding and model-based steganography, already hinted
by~\eqref{eq:rhoH}, in
Section~\ref{sec:hist-pres-lsb}. %

\subsubsection*{Hamming distance}
An even more particular binary case is the one which the support set
of the host is $\mathcal{V}=\{0,1\}$. %
In this case it is quite natural to measure the embedding distortion
using the Hamming distance. Incidentally, the squared Euclidean
distance between two vectors $\mathbf{x},\mathbf{y}\in\{0,1\}^n$
is equal to their Hamming distance or to the Hamming weight
of the watermark, that is to say,
$\delta(\mathbf{y},\mathbf{x})=\|\mathbf{y}-\mathbf{x}\|^2=\|\mathbf{w}\|^2=\omega(\mathbf{w})$. Therefore,
the average degree of host change is completely equivalent to
distortion measures based on the squared Euclidean distance: letting
$\overline{\omega(\mathbf{w})}=\overline{\|\mathbf{w}\|^2}$, from the
previous considerations it now holds that
\begin{equation}
  \label{eq:nu_eq_avw}
  \frac{1}{n}\,\overline{\omega(\mathbf{w})}=\overline{\nu}.
\end{equation}
Also the Hamming weight of~$\mathbf{x}$ is now equal to its squared
norm, $\omega(\mathbf{x})=\|\mathbf{x}\|^2$.  A further identity that
holds true in this case is
$\|\mathbf{x}\|^2=\mathbf{x}^t\mathbf{1}$. As a consequence of these
facts, the theoretical analyses in Sections~\ref{sec:squar-eucl-dist}
and~\ref{sec:average-degree-host} can be particularized, and new
rate-distortion results can be found. Firstly see
that~\eqref{eq:wm_av_2} (or equivalently $n\,\overline{\nu}$) becomes
\begin{equation}
  \overline{\omega(\mathbf{w})}=2\omega(\mathbf{x})\left(1-\frac{1}{n}\omega(\mathbf{x})\right).\label{eq:av_ham}
\end{equation}
Considering~\eqref{eq:nu_eq_avw} and~\eqref{eq:nubounds}, it now holds that
$(1/n)\overline{\omega(\mathbf{w})} \le  1/2$
with equality when $(1/n)\omega(\mathbf{x})=1/2$. On the
other hand,~\eqref{eq:wmax} now takes the form
\begin{equation}
  \omega(\mathbf{w})_\mathrm{max}=2\min(\omega(\mathbf{x}),n-\omega(\mathbf{x})).\label{eq:max_ham} %
\end{equation}
Using~\eqref{eq:av_ham} and~\eqref{eq:max_ham}, the figures of merit
for the embedding distortion~\eqref{eq:pdawr_def}
and~\eqref{eq:ineq_fig_merit} can be put as a function of
$\overline{\omega}(\mathbf{x})\triangleq(1/n)\omega(\mathbf{x})=p_2$
as follows:
\begin{equation}
  \underline{\xi}=\frac{1}{2(1-\overline{\omega}(\mathbf{x}))},\label{eq:xi_av_hamm}
\end{equation}
\begin{equation}
  \xi_\mathrm{min}=\frac{1}{2}\max\left(1,\frac{\overline{\omega}(\mathbf{x})}{1-\overline{\omega}(\mathbf{x})}\right)\label{eq:xi_min_ham}
\end{equation}
whereas
$\underline{\xi}'=1/\left(2\overline{\omega}(\mathbf{x})(1-\overline{\omega}(\mathbf{x}))\right)=1/\overline{\nu}$. The
inequalities in~\eqref{eq:ineq_fig_merit} must still hold;
additionally, we have from~\eqref{eq:nubounds} that
$\underline{\xi}\ge 2\,\overline{\omega}(\mathbf{x})$. Since
$\mathbf{x}$ is in the nonnegative orthant, we can see that
$\xi_\mathrm{min}\ge 1/2$, with equality when
$\overline{\omega}(\mathbf{x})\le 1/2$ (as discussed in
Section~\ref{sec:geometry}). Although~\eqref{eq:xi_min_lb} must still
hold, we can now combine~\eqref{eq:xi_av_hamm}
and~\eqref{eq:xi_min_ham} to obtain the following exact relationship
between these two amounts:
\begin{equation*}
  \xi_\mathrm{min}=\frac{1}{2}\max\left(1,2\underline{\xi}-1\right).\label{eq:xi_min_av_ham}
\end{equation*}

Finally, in the binary Hamming case we have
from~\eqref{eq:up_low_bounds_rho} that $\rho\le h(p_2)$. Therefore we
have an exact relationship between the asymptotic embedding rate
$\rho\doteq h(p_2)=h(\overline{\omega}(\mathbf{x}))$ and the average
embedding distortion~\eqref{eq:nu_eq_avw} since both these amounts
only depend on~$\overline{\omega}(\mathbf{x})$. Thus the
rate-distortion bounds in Section~\ref{sec:rate-dist-bounds}, although
still valid, are asymptotically unnecessary in this case. In any case~\eqref{eq:rd-function} is not useful now, as it can be
greater than one; on the other hand Fano's inequality~\eqref{eq:fano}
now becomes 
\begin{equation}\label{eq:fano2}
  \rho\le h(\overline{\nu}).
\end{equation}
This bound, deduced here for permutation coding with binary base
vector~$\mathbf{x}$, was previously reported to hold for general
binary block codes in~\cite[Theorem
1]{cohen83:nonconstrutive}. Also~\eqref{eq:fano2} can be tightened in
this case using the well-known inequality of the binary entropy
function $h(p_2)\le 2\sqrt{p_2(1-p_2)}$, which allows us to write the
non-asymptotic rate-distortion upper bound
\begin{equation}\label{eq:rd-bound-hamming}
  \rho\le \rho_u''\triangleq \sqrt{2\overline{\nu}},
\end{equation}
which is sharper than~\eqref{eq:fano2} when
$\overline{\nu}\in(0.077,1/2)$. %

\section{Embedding Distortion Control}
\label{sec:embedd-dist-contr}

As argued in Section~\ref{sec:prior-work}, practical universal
steganography requires enforcing similarity between $\mathbf{x}$ and
$\mathbf{y}$ (see Figure~\ref{fig:general}). Nonetheless, a permutation code based on~$\mathbf{x}$
---a fixed input parameter for the encoder--- does not ensure by
itself compliance with some pre-established embedding distortion
constraint, such as for instance a constraint on the maximum value of
$\overline{\nu}$ or on the minimum value
of~$\underline{\xi}$. Critically, we discussed in
Section~\ref{sec:norm-embedd-dist} that $\underline{\xi}$ can be very
low. However similarity between $\mathbf{x}$ and $\mathbf{y}$ can be
increased by restricting the encoding to a judiciously chosen subset of
the set of all permutation codewords. In this section we 
introduce and analyze partitioned permutation coding, which
enables embedding distortion control by means of a similarity
increasing strategy based on the principle just mentioned.

\paragraph{Definitions} A \textit{partitioning} (or \textit{regular
  partitioning}) is defined to be a set of~$p$ index vectors
$\mathcal{U}_n^p=\{\mathbf{u}_1,\mathbf{u}_2,\ldots,\mathbf{u}_p\}$
such that: 1)
$[\mathbf{u}_1^t,\mathbf{u}_2^t,\ldots,\mathbf{u}_p^t]^t\in\mathcal{S}_n$;
and 2) $\mathbf{u}_j={\mathbf{u}^\uparrow_j}$. Consequently, the
lengths $n_1,n_2,\ldots, n_p$ of
$\mathbf{u}_1,\mathbf{u}_2,\ldots,\mathbf{u}_p$ fulfill
$\sum_{j=1}^p n_j=n$. A partitioning $\mathcal{U}_n^p$ is applied to
an $n$-vector~$\mathbf{x}$ by forming $p$
\textit{partitions} %
$\mathbf{x}_1,\mathbf{x}_2,\ldots,\mathbf{x}_p$, which are vectors
formed by elements of $\mathbf{x}$ indexed by the index
vectors. Thus, for $j=1,2,\ldots,p$ the elements of
partition %
$\mathbf{x}_j$ are $(\mathbf{x}_j)_i=(\mathbf{x})_{(\mathbf{u}_j)_i}$
with $i=1,2,\ldots, n_j$. The application of partitioning
$\mathcal{U}_n^p$ to $\mathbf{x}$ is denoted by
$\mathcal{U}_n^p(\mathbf{x})=\{\mathbf{x}_1,\mathbf{x}_2,\ldots,\mathbf{x}_p\}$,
which will also be referred to as a partitioning of~$\mathbf{x}$.

A \textit{support partitioning} is defined to be a partitioning
$\mathcal{U}_q^p$, which is to be applied to the support
$q$-vector~$\mathbf{v}$. A
partitioning of $\mathbf{v}$, i.e.,
$\mathcal{U}_q^p(\mathbf{v})=\{\mathbf{v}_1,\mathbf{v}_2,\ldots,\mathbf{v}_p\}$,
can be used in turn to build a \textit{support-induced partitioning}
of~$\mathbf{x}$, which we denote by
$\mathcal{U}_n^p(\mathcal{U}_q^p,\mathbf{x})=\{\mathbf{x}_1,\mathbf{x}_2,\ldots,\mathbf{x}_p\}$. This
is obtained through the application to~$\mathbf{x}$ of a regular
partitioning $(\mathcal{U}^p_n)^\star$ dependent on
$\mathcal{U}_q^p(\mathbf{v})$ and on~$\mathbf{x}$, whose index vector
$\mathbf{u}^{\star}_j$ contains all indices~$i$
such that $x_i=(\mathbf{v}_j)_k$, where $1\le i\le n$ and $1\le k\le n_j$. Hence
$\mathcal{U}_n^p(\mathcal{U}_q^p,\mathbf{x})=(\mathcal{U}^p_n)^\star(\mathbf{x})$.

The partitions %
in~$\mathcal{U}_q^p(\mathbf{v})$ necessarily coincide with the
supports of the partitions %
in $\mathcal{U}_n^p(\mathcal{U}_q^p,\mathbf{x})$. Therefore the
lengths $q_1,q_2,\ldots,q_p$ of
$\mathbf{v}_1,\mathbf{v}_2,\ldots,\mathbf{v}_p$ fulfill
$\sum_{j=1}^p q_j=q$, not only for $\mathcal{U}_q^p(\mathbf{v})$ (by
the definition of partitioning) but also for the lengths of the
supports of the partitions in
$\mathcal{U}_n^p(\mathcal{U}_q^p,\mathbf{x})$. Note that for a regular
partitioning of $\mathbf{x}$ [i.e., $\mathcal{U}_n^p(\mathbf{x})$] we
have instead that $\sum_{j=1}^p q_j\ge q$, because it is possible that
the same support value $v\in\mathcal{V}$ may be found in more than one
partition %
of~$\mathbf{x}$.

\textit{Example 1}: In order to clarify these concepts, consider
the host $\mathbf{x}=[v_1,v_3,v_2,v_3,v_1,v_3]^t$ with support
$\mathbf{v}=[v_1,v_2,v_3]^t$ and histogram $\mathbf{h}=[2,1,3]^t$
($q=3$). An example of a possible regular partitioning is
$\mathcal{U}_6^2=\{\mathbf{u}_1,\mathbf{u}_2\}$ with index vectors
$\mathbf{u}_1=[2,3,4,5]^t$ and $\mathbf{u}_2=[1,6]^t$ which yields
$\mathcal{U}_6^2(\mathbf{x})=\{\mathbf{x}_1,\mathbf{x}_2\}$ with
partitions %
$\mathbf{x}_1=[v_3,v_2,v_3,v_1]^t$ and $\mathbf{x}_2=[v_1,v_3]^t$.
The supports of these partitions %
are $\mathbf{v}_1=[v_1,v_2,v_3]^t$ ($q_1=3$) and
$\mathbf{v}_2=[v_1,v_3]^t$ ($q_2=2$), and thus $q_1+q_2>q$. 

An example
of a possible support partitioning is
$\mathcal{U}_3^2=\{\mathbf{u}_1',\mathbf{u}_2'\}$ with index vectors
$\mathbf{u}_1'=[1,2]^t$ and $\mathbf{u}_2'=[3]$ which yields
$\mathcal{U}_3^2(\mathbf{v})=\{\mathbf{v}_1',\mathbf{v}_2'\}$ with
partitions %
$\mathbf{v}_1'=[v_1,v_2]^t$ and $\mathbf{v}_2'=[v_3]$. In this case $(\mathcal{U}_6^2)^\star$
contains ${\mathbf{u}_1^\star}'=[1,3,5]^t$ and
${\mathbf{u}_2^\star}'=[2,4,6]^t$, and so
$\mathcal{U}_6^2(\mathcal{U}_3^2,\mathbf{x})=(\mathcal{U}_6^2)^\star(\mathbf{x})=\{\mathbf{x}_1',\mathbf{x}_2'\}$
contains partitions %
$\mathbf{x}_1'=[v_1,v_2,v_1]^t$ and $\mathbf{x}_2'=[v_3,v_3,v_3]^t$.
The lengths of the support partitions $\mathbf{v}_1'$ and
$\mathbf{v}_2'$ (which are also the supports of $\mathbf{x}_1'$ and
$\mathbf{x}_2'$) are %
$q_1'=2$ and $q_2'=1$, and thus $q_1'+q_2'=q$.

\paragraph{Partitioned permutation coding} The previous definitions 
suffice to describe partitioned permutation coding. Encoder and
decoder share a partitioning $\mathcal{U}_n^p$. The encoder
obtains~$\mathcal{U}_n^p(\mathbf{x})$ and then undertakes permutation
coding separately on each of the $p$ partitions; %
 in short, it produces
$\mathbf{y}_j=\Pi_{\boldsymbol\sigma_j}\mathbf{x}_j$ with
$\boldsymbol\sigma_j\in \mathcal{S}_{\mathbf{x}_j}$ for
$j=1,2,\ldots,p$.  In practice, $\mathbf{y}_j$ is produced by
undertaking adaptive arithmetic decoding of $\lfloor \log r_j\rfloor$
bits of the message to be embedded, relying on the histogram
$\mathbf{h}_j$ of~$\mathbf{x}_j$ as described in
Section~\ref{sec:near-optim-embedd}.  The elements of
vector~$\mathbf{y}$ are obtained by piecing together all
$\mathbf{y}_j$ partitions, %
and so
$(\mathbf{y})_{(\mathbf{u}_j)_i}=(\mathbf{y}_j)_i$ with $i=1,2,\ldots,
n_j$, for $j=1,2,\ldots, p$. The vector $\mathbf{y}$ thus obtained
still preserves the histogram of~$\mathbf{x}$ because it stems from
permutations of partitions %
of $\mathbf{x}$, and hence
$\mathbf{y}=\Pi_{\boldsymbol\sigma}\mathbf{x}$ for
some~$\boldsymbol\sigma\in\mathcal{S}_\mathbf{x}$. The decoder obtains
$\mathcal{U}_n^p(\mathbf{y})$ and then, for $j=1,2,\ldots,p$, it
undertakes adaptive arithmetic encoding of partition %
 $\mathbf{y}_j$
relying on its histogram $\mathbf{h}_j$, thus retrieving all~$p$ parts
of the message.

Alternatively, encoder and decoder may share a support
partitioning~$\mathcal{U}_q^p$ instead of a regular
partitioning~$\mathcal{U}_n^p$. In this case the encoder generates
$\mathbf{y}$ from~$\mathcal{U}_n^p(\mathcal{U}_q^p,\mathbf{x})$, and
the decoder retrieves the message in~$\mathbf{y}$
using~$\mathcal{U}_n^p(\mathcal{U}_q^p,\mathbf{y})$.

\subsection{Theoretical analysis of partitioned permutation coding}
\label{sec:theor-analys-with}
Before proceeding, we need to rederive the most relevant results in
Sections~\ref{sec:embedding-rate-deg-eff}
and~\ref{sec:embedd-dist-geom} for partitioned permutation coding with
a generic partitioning~$\mathcal{U}^p_n$. Our computations will
explicitly use
$\mathcal{U}^p_n(\mathbf{x})=\{\mathbf{x}_1,\mathbf{x}_2,\ldots,\mathbf{x}_p\}$,
which is the point of view of the encoder. However, be aware that we
may also evaluate the theoretical expressions that we will obtain
using instead
$\mathcal{U}^p_n(\mathbf{y})=\{\mathbf{y}_1,\mathbf{y}_2,\ldots,\mathbf{y}_p\}$
for any rearrangement~$\mathbf{y}$, which would be the point of view
of the decoder.

Firstly, the number of embeddable messages is 
\begin{equation}
  r=\prod_{j=1}^p r_j, \label{eq:rpart}
\end{equation}
where $r_j={n_j \choose \mathbf{h}_j}$ is the multinomial coefficient
associated to partition~$\mathbf{x}_j$. Hence, the theoretical embedding rate $\rho=(1/n)\log r$
can be expressed as
\begin{equation}
  \rho=\sum_{j=1}^p \frac{n_j}{n} \rho_j, \label{eq:R_partitioning_2}
\end{equation}
where $\rho_j=(1/n_j)\log r_j$ is the embedding rate for the $j$-th partition.

As for the embedding distortion results, firstly see that the average
watermark power for partitioned permutation coding is
$\overline{\|\mathbf{w}\|^2}=(1/r)\sum_{m_1=1}^{r_1}\cdots\sum_{m_p=1}^{r_p}\sum_{j=1}^p\|\mathbf{w}_j^{(m_j)}\|^2$,
where $\mathbf{w}_j^{(m_j)}=\mathbf{y}_j^{(m_j)}-\mathbf{x}_j$. This
amount can be developed as follows
\begin{eqnarray}\label{eq:wm_av_1_partitioning}
  \overline{\|\mathbf{w}\|^2}
  &=&\sum_{j=1}^p\frac{1}{r}\Bigg(\prod_{i=1 \above 0pt i\neq
    j}^{p}r_i\Bigg)\sum_{m_j=1}^{r_j}\|\mathbf{w}_j^{(m_j)}\|^2=\sum_{j=1}^p\overline{\|\mathbf{w}_j\|^2},\nonumber\\
\end{eqnarray}
and hence
\begin{equation}\label{eq:wm_av_2_partition}
  \overline{\|\mathbf{w}\|^2}=2\,\Bigg(\|\mathbf{x}\|^2-\sum_{j=1}^p\frac{1}{n_j}(\mathbf{x}^t_j\mathbf{1})^2\Bigg).
\end{equation}
In parallel to~\eqref{eq:emp_var}, the average watermark power per
host element can be put in terms of the sample variances per partition,
$\sigma^2_{\mathbf{x}_j}=\overline{\|\mathbf{w}_j\|^2}/(2n_j)$, as follows:
\begin{equation}
\frac{1}{n}\overline{\|\mathbf{w}\|^2}=2\sum_{j=1}^p \frac{n_j}{n}\;\sigma^2_{\mathbf{x}_j}.  \label{eq:sample_var_part}
\end{equation}
The maximum watermark power is obtained when $\|\mathbf{w}_j\|^2$ is
maximum for all $j=1,2,\ldots,p$, and then
$\|\mathbf{w}\|^2_\mathrm{max}=\sum_{j=1}^{p}(\|\mathbf{w}_j\|^2)_\mathrm{max}$.
Thus, we have that
\begin{equation}
  \|\mathbf{w}\|^2_\mathrm{max}=2\,\left(\|\mathbf{x}\|^2-\sum_{j=1}^p{{\mathbf{x}^\uparrow_j}}^t{\mathbf{x}^\downarrow_j}\right). \label{eq:wm_max_partition}
\end{equation}
From expressions~\eqref{eq:wm_av_2_partition}
and~\eqref{eq:wm_max_partition} one obtains the figures of merit
$\underline{\xi}$ ($\underline{\xi}'$) and $\xi_\mathrm{min}$ for the
partitioned problem (see definitions in
Section~\ref{sec:norm-embedd-dist}). Importantly,
inequalities~\eqref{eq:ineq_fig_merit} and~(\ref{eq:xi_min_lb}) still
hold with partitioned permutation coding, because
inequalities~\eqref{eq:avmax} and~(\ref{eq:inequalities}) hold for the
average and maximum of $\|\mathbf{w}_j\|^2$, for every
$j=1,2,\ldots,p$.

The number of host changes caused within the $j$-th partition by 
permutation $\boldsymbol\sigma_j\in \mathcal{S}_{\mathbf{x}_j}$ is
computed using a $q_j\times n_j$ matrix $\Lambda_j$ defined just like
$\Lambda$ in Section~\ref{sec:average-degree-host} but using
partitions %
$\mathbf{v}_j$ and~$\mathbf{x}_j$.  Hence, using 
$\Omega_j=\Lambda_j^t\Lambda_j$, for $j=1,2,\ldots,p$, we can write
the average degree of host change for partitioned permutation coding
as
$\overline{\nu}=(1/r)\sum_{\boldsymbol\sigma_1\in\mathcal{S}_{\mathbf{x}_1}}
\cdots \sum_{\boldsymbol\sigma_p\in\mathcal{S}_{\mathbf{x}_p}}
(1/n)\left( \sum_{j=1}^p n_j-\tr
  \Omega_j\Pi_{\boldsymbol\sigma_j}\right)$. This expression can be
developed  like~\eqref{eq:wm_av_1_partitioning} to yield
\begin{eqnarray}
  \label{eq:adhc_part}
  \overline{\nu}
  &=&\sum_{j=1}^p\frac{1}{r}\Bigg(\prod_{i=1 \above 0pt i\neq
    j}^{p}r_i\Bigg)
  \sum_{\boldsymbol\sigma_j\in\mathcal{S}_{\mathbf{x}_j}}\frac{1}{n}\left(n_j-\tr
    \Omega_j\Pi_{\boldsymbol\sigma_j}\right)=\sum_{j=1}^p\frac{n_j}{n}\;\overline{\nu}_j,\nonumber\\
\end{eqnarray}
where $\overline{\nu}_j=1-(\|\mathbf{h}_j\|/n_j)^2=1-\|\mathbf{p}_j\|^2$.  

Finally, the average embedding efficiency for partitioned
permutation coding is
\begin{equation}
  \label{eq:eepart}
  \overline{\varepsilon}=\frac{1}{r}\sum_{\boldsymbol\sigma_1, \ldots,
    \boldsymbol\sigma_p\backslash\boldsymbol\sigma_0}\frac{\log r}{\sum_{j=1}^p
    n_j-\tr \Omega_j\Pi_{\boldsymbol\sigma_j}}.
\end{equation}
The first summation in~\eqref{eq:eepart} ranges over all
$\boldsymbol\sigma_j\in\mathcal{S}_{\mathbf{x}_j}$, for
$j=1,2,\ldots,p$, bar the case
$\boldsymbol\sigma_0\in\mathcal{S}_\mathbf{x}$ in which
$\mathbf{y}=\mathbf{x}$.  Using the same bounding strategy as in
Section~\ref{sec:embedding-efficiency}, the lower bound
$\overline{\varepsilon}_l$ to~\eqref{eq:eepart} is
again~\eqref{eq:average_epsilon}, where $r$, $\rho$ and
$\overline{\nu}$ are now given by \eqref{eq:rpart},
\eqref{eq:R_partitioning_2} and~\eqref{eq:adhc_part}, respectively. It
can also be verified by applying~\eqref{eq:rd-nu2} to each $\rho_j$
that~\eqref{eq:ineq_vareps} also holds for partitioned permutation
coding.

We verify next that the rate-distortion bounds in
Section~\ref{sec:rate-dist-bounds} still hold.
Applying~\eqref{eq:rd-function} individually to the partition rates
$\rho_j$ in~\eqref{eq:R_partitioning_2}, and then using the concavity
of $\log(\cdot)$ and Jensen's inequality, we have that
\begin{eqnarray}
  \rho&< &\sum_{j=1}^p \frac{n_j}{n}\; \frac{1}{2} \log\left(2\pi e\;
    \left(\sigma^2_{\mathbf{x}_j}+\frac{1}{12}\right)\right)\label{eq:rd-function-part-1}\\
  &\le&\frac{1}{2}\log\left(2\pi e\;\sum_{j=1}^p \frac{n_j}{n} 
    \left(\sigma^2_{\mathbf{x}_j}+\frac{1}{12}\right)\right).   \label{eq:rd-function-part}
\end{eqnarray}
Therefore, using~\eqref{eq:sample_var_part}
in~\eqref{eq:rd-function-part} we recover~\eqref{eq:rd-function}.
Note that the inequality leading to~\eqref{eq:rd-function-part} (Jensen's inequality) is met
with equality if and only if
$\sigma^2_{\mathbf{x}_j}=\overline{\|\mathbf{w}\|^2}/(2n)$ for all
$j=1,2,\ldots, p$. Similarly, applying~\eqref{eq:fano} individually to
each $\rho_j$ in~\eqref{eq:R_partitioning_2} and then using the
concavity of $h(\overline{\nu}_j)+\log(q_j-1)\,\overline{\nu}_j$ on
$\overline{\nu}_j$, $\log q_j\le \log q$, Jensen's inequality
and~\eqref{eq:adhc_part}, we see that~\eqref{eq:fano} still holds with
partitioning.

We finally prove that the rate-distortion lower
bounds~\eqref{eq:bound-rate-dhc} and~\eqref{eq:rd-nu2} also hold for
the class of support-induced
partitionings. Applying~\eqref{eq:bound-rate-dhc} individually to the
partition rates $\rho_j$ in~\eqref{eq:R_partitioning_2}, and then
using the convexity of $-\log(\cdot)$ and Jensen's inequality, we have
that
\begin{eqnarray}
  \rho&\ge &\sum_{j=1}^p \frac{n_j}{n}\;\left(-\log\left(1-\overline{\nu}_j\right)-\zeta^{(q_j,n_j)}\right)\label{eq:bound-rate-dhc-part-previous}\\
  &\ge&-\log\left(\sum_{j=1}^p \frac{n_j}{n}
    (1-\overline{\nu}_j)\right)-\frac{q}{n}\log\left(\sum_{j=1}^p\frac{q_j}{q} n_j+1\right),\nonumber\\   \label{eq:bound-rate-dhc-part}
\end{eqnarray}
because $\sum_{j=1}^p (q_j/q)=1$ for support-induced partitionings.
Therefore, using~\eqref{eq:adhc_part} and $\sum_{j=1}^p (q_j/q) n_j\le
n$ in~\eqref{eq:bound-rate-dhc-part} we
recover~\eqref{eq:bound-rate-dhc}. Following the same steps as
in~\eqref{eq:bound-rate-dhc-part-previous}
and~\eqref{eq:bound-rate-dhc-part}, it is seen
that~\eqref{eq:rd-nu2} still holds for support-induced partitionings.

Lastly, we describe the invariance property of support partitionings
with respect to the theoretical analysis in this Section~\ref{sec:theor-analys-with}, which will be
seen to be key for the optimization of partitioned permutation coding
in Section~\ref{sec:part-select}. As
discussed therein, this property allows the decoder to determine the
partition dynamically chosen by the encoder in order to meet a
distortion constraint, without the need for a side channel for the
encoder to communicate its choice to the decoder.

\paragraph*{Property (Invariance  of support partitionings)} Given a
support partitioning~$\mathcal{U}_q^p$ and any arbitrary rearrangement
$\mathbf{y}$ of the host~$\mathbf{x}$, all theoretical predictions for
partitioned permutation coding are identical when using either the
support-induced partitioning
$\mathcal{U}_n^p(\mathcal{U}_q^p,\mathbf{x})$ or the support-induced
partitioning $\mathcal{U}_n^p(\mathcal{U}_q^p,\mathbf{y})$. In
particular, there is always agreement between both cases
on~\eqref{eq:R_partitioning_2},~\eqref{eq:wm_av_2_partition},~\eqref{eq:wm_max_partition}
and~\eqref{eq:adhc_part}, and consequently also on
$\overline{\varepsilon}_l$ and all bounds
(\ref{eq:rd-function}--\ref{eq:rd-nu2}) which we have seen always hold
for support-induced partitionings. This behavior is due to the fact
that, since the support partitions
$\mathbf{v}_1,\mathbf{v}_2,\ldots,\mathbf{v}_p$ must be identical for
any two partitionings induced by~$\mathcal{U}_q^p$, then
$\mathbf{y}_j$ must always be a rearrangement of $\mathbf{x}_j$ for
$j=1,2,\ldots,p$. Seeing that all
expressions~(\ref{eq:rpart}--\ref{eq:adhc_part}) %
are independent of any particular rearrangement of the values within
any of the $p$ partitions, and that all rate-distortion bounds in
Section~\ref{sec:rate-dist-bounds} hold for support-induced
partitionings, then it follows that the evaluation of every
theoretical expression using
$\mathcal{U}_n^p(\mathcal{U}_q^p,\mathbf{x})$ or
$\mathcal{U}_n^p(\mathcal{U}_q^p,\mathbf{y})$ coincides.

We must remark that the invariance property does not hold in general
for a regular partitioning~$\mathcal{U}_n^p$. In this case,
theoretical predictions using~$\mathcal{U}_n^p(\mathbf{x})$ and
$\mathcal{U}_n^p(\mathbf{y})$ only necessarily coincide in the
particular case in which~$\mathbf{y}$ stems from applying partitioned
permutation coding to~$\mathbf{x}$ using~$\mathcal{U}_n^p$ (as in this
case $\mathbf{y}_j$ must be a rearrangement of $\mathbf{x}_j$ for
$j=1,2,\ldots,p$).

\textit{Example 2}: Consider again the host $\mathbf{x}$ and the
partitionings $\mathcal{U}_6^2$ and $\mathcal{U}_3^2$ given in
Example~1. In order to numerically illustrate the invariance property
of support partitionings, assume that $\mathbf{v}=[1,2,3]^t$, and
consider the rearrangement $\mathbf{y}=[2, 1, 3, 3, 3, 1]^t$ of
$\mathbf{x}=[1,3,2,3,1,3]^t$. If the theoreticals are computed using
$\mathcal{U}_6^2(\mathbf{x})=\{[3,2,3,1]^t,[1,3]^t\}$, the main
results are $\rho=4.58$, $\overline{\|\mathbf{w}\|^2}=9.5$,
$\|\mathbf{w}\|^2_\text{max}=18$, $\overline{\varepsilon}_l=1$, and
$\overline{\nu}=0.59$; however, the values obtained using
$\mathcal{U}_6^2(\mathbf{y})=\{[1,3,3,3]^t,[2,1]^t\}$ are different:
$\rho=3$, $\overline{\|\mathbf{w}\|^2}=7$,
$\|\mathbf{w}\|^2_\text{max}=10$, $\overline{\varepsilon}_l=0.92$, and
$\overline{\nu}=0.42$.  As $\mathbf{y}$ does not stem from
$\mathcal{U}_6^2(\mathbf{x})$, no invariance is guaranteed.  However,
even though $\mathbf{y}$ does not stem from
$\mathcal{U}_6^2(\mathcal{U}_3^2,\mathbf{x})$ either, the results are
identical for the support-induced partitionings
$\mathcal{U}_6^2(\mathcal{U}_3^2,\mathbf{x})=\{[1,2,1]^t,[3,3,3]^t\}$
and
$\mathcal{U}_6^2(\mathcal{U}_3^2,\mathbf{y})=\{[2,1,1]^t,[3,3,3]^t\}$:
$\rho=1.58$, $\overline{\|\mathbf{w}\|^2}=1.33$,
$\|\mathbf{w}\|^2_\text{max}=2$, $\overline{\varepsilon}_l=0.19$, and
$\overline{\nu}=0.23$ (i.e., we observe the invariance property of
support partitionings). Notice that in $\mathcal{U}_6^2(\mathbf{x})$
and $\mathcal{U}_6^2(\mathbf{y})$ the partitions are not
rearrangements of each other, whereas in
$\mathcal{U}_6^2(\mathcal{U}_3^2,\mathbf{x})$ and
$\mathcal{U}_6^2(\mathcal{U}_3^2,\mathbf{y})$ they are (in this toy
example, the second partition is identical in both cases).

\subsection{Static and adaptive partitioning }
\label{sec:part-select}

As we have seen, encoder and decoder can always implement partitioned
permutation coding by sharing a partitioning~$\mathcal{U}_n^p$ (or
else a support partitioning $\mathcal{U}_q^p$). This strategy will
lead to an embedding distortion which may or may not comply with a
given constraint. Since in this case the shared partitioning is
predetermined, we will call this strategy \textit{static}
partitioning. %

For partitioned permutation coding to comply with a distortion
constraint, \textit{adaptive} partitioning is required. Adaptive
partitioning is an optimization problem where a
partitioning~$\mathcal{U}_n^p$ must be chosen for host $\mathbf{x}$
such that the constraint is met and the embedding rate is
maximized. In the following we will only consider a maximum constraint
on $\overline{\|\mathbf{w}\|^2}$ (equivalently, a minimum constraint
on $\underline{\xi}$ or on $\underline{\xi}'$).  The reason is
twofold: 1) the average watermark power asymptotically approximates
the watermark power associated to any random rearrangement
of~$\mathbf{x}$ (Section~\ref{sec:asymptotic}), and it also bounds the
maximum power through~\eqref{eq:xi_min_lb}; and 2) from~\eqref{eq:nuw}
and~\eqref{eq:bound_nu} one sees that a maximum constraint on
$\overline{\|\mathbf{w}\|^2}$ implies a maximum constraint
on~$\overline{\nu}$, whereas $\overline{\|\mathbf{w}\|^2}$ is more
convenient to assess the suitability of a partitioning via the
rate-distortion upper bound~\eqref{eq:rd-function}. With these
considerations in mind, adaptive partitioning involves solving the
optimization problem
\begin{equation}
  \label{eq:rho_max_part}
  \rho^*=\max_{\mathcal{U}_n^p(\mathbf{x}):\; \underline{\xi}\,\ge\kappa} \rho,
\end{equation}
where $\rho$ and $\underline{\xi}$ are the theoretical embedding rate
and the theoretical embedding distortion corresponding to
$\mathcal{U}_n^p(\mathbf{x})$, and $\kappa$ is the constraint. The
optimization in~\eqref{eq:rho_max_part} is of a combinatorial
nature. If the encoder were able to solve~\eqref{eq:rho_max_part}, it
would then produce $\mathbf{y}$ using the optimum
$(\mathcal{U}_n^p)^*$ found. The decoder would then need the very same
partitioning before it could proceed to decode the message. The
simplest possibility would be sending $(\mathcal{U}_n^p)^*$ to the
decoder through a side channel. However this style of
\textit{nonblind} adaptive partitioning is not permissible, since
using a side channel defeats the purpose of steganography.

Therefore \textit{blind} adaptive partitioning is the only way
forward.  This requires that the decoder be able to obtain
$(\mathcal{U}_n^p)^*$ without knowledge of~$\mathbf{x}$, i.e.,
by undertaking the maximization in~\eqref{eq:rho_max_part} on
$\mathcal{U}_n^p(\mathbf{y})$, rather than on
$\mathcal{U}_n^p(\mathbf{x})$. This would be a legitimate strategy
provided that the same~$\rho^*$ would correspond to a
unique~$(\mathcal{U}_n^p)^*$ in the distinct optimization problems of 
encoder and decoder, but there are no guarantees of this happening in
general.  This issue can be circumvented by constraining the potential solutions to be in the class of support-induced partitionings. Due to
the invariance property of support partitionings, both parties will
separately agree on the theoretical performance associated to every
possible support-induced partitioning, and thus, in theory, they will be able
find the optimum partitioning(s) within this class. In
order to break any ties, encoder and decoder can share a sequence of
all support partitionings $\{\mathcal{U}_q^p\}$ and choose, for
instance, the first optimum in the
sequence. %

Apart from the fact that the class of support-induced partitionings may
not contain the optimum in~\eqref{eq:rho_max_part}, the major issue
with this strategy %
is the size of the space to be searched. The total number of support
partitionings in~$\{\mathcal{U}_q^p\}$ is the Bell number
$\beta_q\triangleq\sum_{p=1}^q S(q,p)$, %
as, by definition,
the Stirling number of the second kind $S(q,p)$ %
gives the
number of partitionings of a set of cardinality $q$ into $p$ nonempty
partitions. For example, for a host $\mathbf{x}$ represented
using~$b=8$ bits/element, $q=256$ and we have $\beta_{256}\sim
10^{373}$ possible support partitionings\footnote{When $q=2$ there is
  only one support partitioning with nonzero rate, and hence blind
  adaptive partitioning is not possible for binary
  hosts. %
}. The problem can be somewhat relaxed if we only consider support
partitionings with connected partitions (that is to say, index vectors
with adjacent indices), which makes intuitive sense in terms of
distortion minimization. The total number of such partitionings is
$\beta_q'\triangleq\sum_{p=1}^q {q-1 \choose p-1}=2^{q-1}$. This can
be seen from the bijection between the support partitionings with $p$
nonempty connected partitions and the binary strings of length $q-1$
and Hamming weight $p-1$, which implies that there are ${q-1 \choose
  p-1}$ such partitionings. Using again $q=256$ we now have
$\beta_{256}'\sim 10^{76}$ possible support partitionings, a %
noticeably smaller number than $\beta_{256}$ but still forbidding.

\subsubsection{Practical blind adaptive partitioning}
\label{sec:pract-select-part}
We will see next that a suboptimal solution to the problem of blind
adaptive partitioning is feasible. The key to an implementable
strategy must be a short but representative sequence of support
partitionings, in order to avoid evaluating the full sequence
$\{\mathcal{U}_q^p\}$ ---which typically is prohibitively long. A
necessary property of partitionings that are candidates to
solve~\eqref{eq:rho_max_part}, and which therefore should be part of
the aforementioned short sequence, can be deduced from the analysis of
the rate-distortion bound~\eqref{eq:rd-function} for partitioned
permutation coding. Observe that this upper bound is sharpest when
Jensen's inequality in~\eqref{eq:rd-function-part} holds with
equality, which happens when the sample variance per partition is
constant, i.e., when $\sigma^2_{\mathbf{x}_j}=\sigma^2_{\mathbf{x}}$
for all $j=1,2,\ldots,p$. Therefore, for a given embedding distortion
constraint, the embedding rate of a partitioning with uneven sample
variance per partition can only approach the upper
bound~\eqref{eq:rd-function-part-1} but
not~\eqref{eq:rd-function-part}. Thus constancy of the intrapartition
sample variance is a necessary ---but not sufficient--- condition for
a partitioning to induce an embedding rate as close as possible
to the upper bound~$\rho_u$ (which, we recall, is independent of any partitioning).

As we will verify in Section~\ref{sec:results}, for a low embedding
distortion constraint (i.e., for a high $\kappa$) the necessary
condition above can be approximated by means of support partitionings
which uniformly divide~$\mathbf{v}$ into~$p$ connected partitions,
which gives a sequence of only~$q$ support partitionings.  This
sequence is denoted
as~$\{\widetilde{\mathcal{U}}_q^{p}\}_{p=1}^{q}$. The $p$-th support
partitioning $\widetilde{\mathcal{U}}_q^{p}$ is obtained by defining
the centroids $c_j\triangleq v_1+ (j-1/2)(v_q-v_1)/p$ for
$j=1,2,\ldots,p$, and then, for $k=1,2,\ldots, q$, by assigning
index~$k$ to index vector $\mathbf{u}_t$ such that
$t=\argmin_{j\in\{1,2,\ldots,p\}} |v_k-c_j|$.  The encoder finds $p$
between $1$ and $q$ such that
$\mathcal{U}_n^p(\widetilde{\mathcal{U}}_q^{p},\mathbf{x})$ complies
with $\underline{\xi}\ge \kappa$ and maximizes $\rho$, and then uses
this support-induced partitioning to produce~$\mathbf{y}$. The decoder
finds~$p$ between $1$ and~$q$ such that
$\mathcal{U}_n^p(\widetilde{\mathcal{U}}_q^{p},\mathbf{y})$ complies
with $\underline{\xi}\ge \kappa$ and maximizes $\rho$, and then uses
this support-induced partitioning to decode~$\mathbf{y}$. In case of
ties, the first complying partitioning in the sequence (for example)
is chosen by both parties.  With respect to the performance of this
approach, the maximum
$\overline{\nu}=\sum_{j=1}^p(n_j/n)(1-1/q_j)\approx 1-p/q$ is achieved
when there is intrapartition uniformity, which turn maximizes the
lower bounds~\eqref{eq:bound-rate-dhc} and~\eqref{eq:rd-nu2}. Now,
intrapartition uniformity is approximated by
$\widetilde{\mathcal{U}}_q^p$ as~$p$ increases, and in this case,
which usually corresponds to large~$\kappa$, \eqref{eq:bound-rate-dhc}
and~\eqref{eq:rd-nu2} close in on~\eqref{eq:rd-function}.

For the reasons discussed above (approximate fulfillment of necessary
condition for approaching upper bound and approximate maximization of
lower bounds for large $\kappa$), the strategy described in
this section can achieve embedding rates reasonably close to the upper
bound~\eqref{eq:rd-function}, as it will be empirically verified in
Section~\ref{sec:results}. It must be remarked
that~\eqref{eq:rd-function} is not attainable with equality.

\subsubsection{A special static partitioning}\label{sec:hist-pres-lsb} 
Leaving behind blind adaptive partitioning, we conclude this section
by analyzing a special static partitioning strategy of particular
interest. The reason is three-fold: a) this special static partitioning leads
to a rate-distortion pair approximately independent of~$\mathbf{x}$
for a large class of hosts; b) it allows to settle a long-standing
problem of LSB-like steganographic algorithms discussed in more detail
below; and c) it affords a fair comparison between partitioned
permutation coding and model-based steganography.

We will assume without loss of generality that the support of
$\mathbf{x}$ contains all consecutive values between
$v_1\leftarrow 2\lfloor v_1/2\rfloor$ and
$v_q\leftarrow 2\lfloor v_q/2\rfloor+1$ (by including, if required,
additional values in the support of~$\mathbf{x}$ with zero-valued
entries in their corresponding histogram positions), and so we may
also assume that~$q$ is even.  With these assumptions, the special
static support partitioning that we will be analyzing is defined
as~$\mathcal{U}_q^{q/2} \triangleq
\{[1,2]^t,[3,4]^t,\ldots,[q-1,q]^t\}$,
and therefore $\mathcal{U}_q^{q/2}(\mathbf{v})$ is formed by $p=q/2$
partitions each containing two adjacent values of the
support~$\mathbf{v}$. Equivalently, every partition~$\mathbf{v}_j$ in
$\mathcal{U}_q^{q/2}(\mathbf{v})$ only contains the pair of adjacent
histogram bins $(\mathbf{v}_j)_1$ and
$(\mathbf{v}_j)_2=(\mathbf{v}_j)_1+1$. Assuming that the
histogram~$\mathbf{h}$ of $\mathbf{x}$ varies slowly, we can make the
approximation $(\mathbf{h}_j)_1\approx (\mathbf{h}_j)_2$, or
$\mathbf{h}_j\approx (n_j/2)\mathbf{1}$, for the histogram of each
partition~$\mathbf{x}_j$ in 
$\mathcal{U}_n^{q/2}(\mathcal{U}_q^{q/2},\mathbf{x})$. %
This is equivalent to saying that there is approximate intrapartition
uniformity, and also that the intrapartition variances are
approximately constant. To see this last point, it is convenient to
rewrite the intrapartition variance in terms of $\mathbf{h}_j$ and
$\mathbf{v}_j$ as follows: $\sigma_{\mathbf{x}_j}^2=
(1/n_j)(\mathbf{v}_j^t\diag(\mathbf{h}_j)\mathbf{v}_j-(1/n_j)(\mathbf{v}_j^t\mathbf{h}_j)^2)$.
Using the approximation of $\mathbf{h}_j$ in this expression we get
\begin{eqnarray}
  \sigma_{\mathbf{x}_j}^2
  &\approx&\frac{1}{2}\left(\|\mathbf{v}_j\|^2-\frac{1}{2}(\mathbf{v}_j^t\mathbf{1})^2\right)\nonumber\\
  &=&\frac{1}{4}\left((\mathbf{v}_j)_2-(\mathbf{v}_j)_1\right)^2=\frac{1}{4}.  \label{eq:intrappart}
\end{eqnarray}
From~\eqref{eq:wm_av_2_partition} and~\eqref{eq:intrappart} we
have that the average embedding distortion per host element can be therefore
approximated as
\begin{eqnarray}
  \frac{1}{n}\overline{\|\mathbf{w}\|^2}&\approx&\frac{1}{2}.
  \label{eq:wm_av_2_partition_lsb}
\end{eqnarray}
Using next~\eqref{eq:wm_av_2_partition_lsb} in upper
bound~\eqref{eq:rd-function} we obtain
$\rho_u\approx\frac{1}{2}\log\left(\frac{2}{3}\pi
  e\right)=1.2546$. Also since $\mathbf{p}_j\approx(1/2)\mathbf{1}$
then from~\eqref{eq:adhc_part} we have that $\overline{\nu}\approx1/2$, and the lower
bounds~\eqref{eq:bound-rate-dhc} and~\eqref{eq:rd-nu2} become
$\rho_l'=\rho_l''~\shortstack[c]{\vspace*{-.05cm}.\\$\approx$}~1$. On
the other hand, assuming that $n_j$ is even for simplicity, the embedding
rate~\eqref{eq:R_partitioning_2} is approximated as follows:
\begin{equation}
  \rho\approx \sum_{j=1}^p\frac{n_j}{n} \left(\frac{1}{n_j}\log
    {n_j\choose \frac{n_j}{2}}\right)\approx 1\text{ bit/host element},\label{eq:rho_partition_lsb}
\end{equation}
where the second approximation in~\eqref{eq:rho_partition_lsb} assumes
that $n_j$
is large for all $j=1,2,\ldots,p$
and uses Stirling's formula. Therefore both the embedding rate and its
lower bounds are approximately $20\%$
below the upper bound for the static support partitioning
$\mathcal{U}_q^{q/2}$. Even if this looks like a modest achievement, bear in
mind that approximations~\eqref{eq:wm_av_2_partition_lsb}
and~\eqref{eq:rho_partition_lsb} hold for any host for which the
histogram assumption is valid, whereas~$\rho_u$
can only be approached (but not attained) by a host whose histogram
resembles a Gaussian distribution.

More importantly, the analysis above shows that partitioned permutation
coding relying on $\mathcal{U}_q^{q/2}$ implements histogram-preserving steganography with
approximately the same embedding distortion and embedding rate
---namely,~\eqref{eq:wm_av_2_partition_lsb}
and~\eqref{eq:rho_partition_lsb}--- as (non histogram-preserving) LSB
steganography. This allows to finally settle a long-standing practical
problem of LSB-like steganographic algorithms: some of them are able
to preserve the histogram at the cost of decreasing the rate (like the
techniques mentioned in Section~\ref{sec:prior-work}), while others
are able to preserve the rate-distortion of LSB steganography at the
cost of only smoothing the histogram artifacts caused by the LSB
method (in particular the extensively studied LSB matching algorithm, also
known as $\pm 1$ steganography~\cite{sharp01:lsb}).

Finally, the static support partitioning strategy in this section also
parallels the quantizer step size~2 embedding distortion constraint in
model-based steganography~\cite{sallee03:model}\footnote{A minor
  difference is that zero-valued entries in $\mathbf{x}$ are untouched
  in model-based steganography (in this algorithm, $\mathbf{x}$
  contains values from a single frequency in the quantized block DCT
  domain). The exact parallel is achieved by using a support
  partitioning having one partition containing only zero, plus
  partitions with two adjacent values for the rest of the
  support~$\mathbf{v}$.}, the first algorithm to use arithmetic coding
to flesh out the duality between steganography and
compression. Considering the algorithm in
Section~\ref{sec:near-optim-embedd} and partitioned permutation coding
with the static support partitioning $\mathcal{U}_q^{q/2}$, one can
see that the defining difference of partitioned permutation coding
with respect to model-based steganography is ---despite their
seemingly unrelated starting points--- the use of adaptive arithmetic
coding with an empirical model of the host, rather than nonadaptive
arithmetic coding with a theoretical model of the host. In the
conditions in this section both algorithms have similar
rate-distortion features, but only partitioned permutation coding is
histogram-preserving and thus perfect for memoryless signals.

\subsection{Considerations on the achievable performance}
\label{sec:part-side-inform}
In order to discuss why partitioned permutation coding can aim at
performing close to $\rho_u$ for a distortion constraint, let us
particularize Gel'fand and Pinsker's result for the achievable rate of
a communications system with noncausal side information at the
encoder~\cite{Gelfand} to the problem addressed in this paper. Since
the channel is noise-free, this result is
\begin{equation}\label{eq:gp_or}
  \rho^*_\text{gp}\triangleq \max_{p(y,u'\vert x)} \left\{\I(Y;U')-\I(U';X)\right\} \text{ bits/host element},
\end{equation}
where the discrete random variables $Y$ and $X$, both with
support~$\mathcal{V}$, represent the information-carrying signal and
the host, respectively, and $U'$ is an auxiliary discrete random
variable whose support has cardinality
$|\mathcal{U}'|\le 2|\mathcal{V}|+1=2q+1$. In our case, the
maximization in~\eqref{eq:gp_or} must take into account two
constraints: 1)
$n\sigma_X^2/\E\left\{\|\boldsymbol{Y}-\boldsymbol{X}\|^2\right\}\ge
\kappa$
(cf. the constraint in~\eqref{eq:rho_max_part}); and 2)
$p(Y=v)=p(X=v)$ for all $v\in\mathcal{V}$ (perfect
steganography). Barron et al.~\cite{Barron03} showed that, under the
first constraint, $Y$ can be taken to be a deterministic function of
$X$ and $U'$, i.e., $Y=g(X,U')$. %
The second constraint allows us to develop the functional
in~\eqref{eq:gp_or} as follows:
\begin{eqnarray}
  \I(Y;U')-\I(U';X)&=&\H(X\vert U')-\H(Y\vert U')   \label{eq:gp_eq}\\
  &\le& \H(X\vert U'),   \label{eq:gp}
\end{eqnarray}
where~\eqref{eq:gp_eq} is because $\H(Y)=\H(X)$ and~\eqref{eq:gp}
because discrete entropy is nonnegative. Equality is achieved
in~\eqref{eq:gp} when $H(Y|U')=0$, which happens when $Y=g(U')$, that
is, when~$Y$ is only a function of $U'$. In this case, in order to
determine $\rho^*_\text{gp}$ it is sufficient to maximize~\eqref{eq:gp}
under the first constraint over
all~$p(u'|x)$. %

Let us next recast the embedding rate~\eqref{eq:R_partitioning_2} of
partitioned permutation coding in information-theoretical terms. As
in~\eqref{eq:rhoH}, we can make the approximation $\rho_j\approx
\H(X\vert U=j)$, where~$U$ is a random variable with support
$\{1,2,\ldots,p\}$ and pmf~$p(U=j)= n_j/n$, which reflects the
probability that~$X$ belongs to the $j$-th partition, and $X|(U=j)$ is
a random variable with support $\mathbf{v}_j$ and pmf
$\mathbf{p}_j=(1/n_j)\mathbf{h}_j$. Thus,~\eqref{eq:R_partitioning_2}
can be approximated as
\begin{equation}
  \rho\approx\sum_{j=1}^p p(U=j)\; \H(X\vert U=j)=\H(X\vert U). \label{eq:R_partitioning_2_hxu}
\end{equation}
This approximation of $\rho$, which is asymptotically exact, has the
same form as the upper bound~\eqref{eq:gp} on Gel'fand and Pinsker's
maximization functional. The partitioning variable~$U$
in~\eqref{eq:R_partitioning_2_hxu} is also analogous to the auxiliary
variable $U'$ in~\eqref{eq:gp}: in the same way that $U'$ determines
$Y$ when equality holds in~\eqref{eq:gp}, the partitioning variable
$U$ suffices to determine~$Y$. For these reasons, the constrained
maximization problem~\eqref{eq:rho_max_part} can be seen as analogous
of the constrained maximization of~\eqref{eq:gp_or}.

\section{Empirical Results}
\label{sec:results}
In this section we verify the correspondence between empirical and
theoretical results. Each empirical value corresponds to one single
message~$m$ drawn uniformly at random from
$\{0,1,\ldots,2^{n\rho_\mathrm{emp}}\}$, where the empirical embedding
rate is $\rho_\textrm{emp}=(1/n)\sum_{j=1}^p \lfloor\log
\tilde{r}_j\rfloor$ bits/host element for a partitioning~$\mathcal{U}_n^p$ (or $\mathcal{U}_q^p$). The $\log \tilde{r}_j$
amounts are lower bounds of $\log r_j$ computed using Robbins'
sharpening of Stirling's formula~\cite{robbins55:_remark}
\begin{equation}
  \sqrt{2\pi z}(z/e)^z e^{(12
    z+1)^{-1}} < z!  <\sqrt{2\pi z}(z/e)^z e^{(12
    z)^{-1}}.\label{eq:robbins}
\end{equation}
The lower bound in~\eqref{eq:robbins} is applied to the factorial in
the numerator of the multinomial coefficient $r_j$, and the upper
bound in~\eqref{eq:robbins} to each of the factorials in its
denominator. Observe that it is essential that $r_j$ is accurately
approximated but never overestimated, because the $j$-th partition
cannot convey more messages than rearrangements of~$\mathbf{x}_j$. The
lower bound in~\eqref{eq:up_low_bounds_rho}, although convenient from
a theoretical point of view, is less sharp than the one obtained
using~\eqref{eq:robbins}. As for the theoretical
rate~\eqref{eq:R_partitioning_2}, $\log r_j$ is bounded by applying
the upper bound in~\eqref{eq:up_low_bounds_rho} to~$\rho_j$. For small
factorials one may drop these bounding strategies and exactly compute
$\tilde{r}_j=r_j$.

\begin{figure}[t!]
  \centering
  \includegraphics[height=6.5cm]{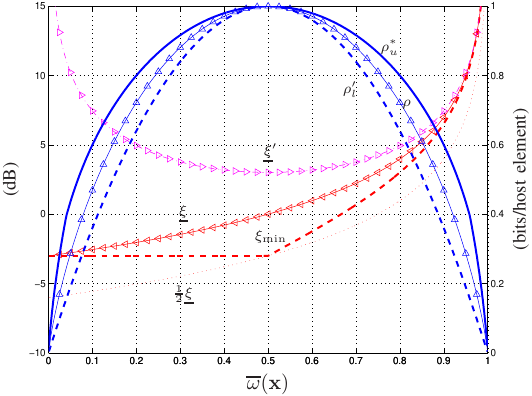}
  \caption{Performance of unpartitioned permutation coding for a host
    with support $\mathcal{V}=\{0,1\}$ and $n=10^6$ as a function of
    its Hamming weight per symbol. Lines are theoretical results and
    symbols are empirical measurements. Rate units are on the
    right-hand side of the plot, whereas distortion units are on the
    left-hand side;  $\rho_u^*=\min(\rho_u',\rho_u'')$.}
  \label{fig:hamming}
\end{figure}

The information-carrying vector $\mathbf{y}=e(\mathbf{x},m)$ is then
produced as described in
Sections~\ref{sec:embedd-dist-contr} and~\ref{sec:near-optim-embedd}. In
every case it is verified that $\mathbf{y}$ preserves the histogram of
$\mathbf{x}$, and that the decoder retrieves the message without
error, i.e., $d(\mathbf{y})=m$. The following empirical amounts are
computed from $\mathbf{w}=\mathbf{y}-\mathbf{x}$:
$\xi_\textrm{emp}=\|\mathbf{x}\|^2/\|\mathbf{w}\|^2$ (empirical host
to watermark power ratio), $\xi_\textrm{emp}'= n
(2^b-1)^2/\|\mathbf{w}\|^2$ (empirical peak host to watermark power
ratio), $\nu_\textrm{emp}=(1/n)\sum_{i=1}^n\ind{w_i\neq 0}$ (empirical
degree of host change) and $\varepsilon_\textrm{emp}=
\rho_\textrm{emp}/\nu_\textrm{emp}$ bits/host element change
(empirical embedding efficiency).

The results in Figure~\ref{fig:hamming} are for unpartitioned
permutation coding in the binary Hamming case. As discussed in
Section~\ref{sec:binary-case}, in this case the asymptotic
rate-distortion performance only depends on the normalized Hamming
weight of the host, and consequently it is particularly easy to
visualize. In the figure, the lower rate-distortion bound is the
asymptotic result $\rho_l'\doteq 2\overline{\nu}$, which is tighter
than $\rho_l$ because from~\eqref{eq:nubounds} we have that
$\overline{\nu}\le 1/2$ for a binary host. The upper bound shown is
the minimum of~\eqref{eq:fano2} and~\eqref{eq:rd-bound-hamming}. The
theoretical rates match the empirical rates accurately, just
because~$r$ is large and the bounds~\eqref{eq:robbins} are
asymptotically tight. It is more interesting to see that the empirical
distortion results ---which correspond to single  watermarks
(i.e., they are one-shot results and not averages)--- accurately match
the theoretical predictions involving averages, that is,
$\underline{\xi}$ and $\underline{\xi}'$. We showed in
Section~\ref{sec:asymptotic} that this asymptotic behavior of the
embedding distortion of permutation coding for large~$n$ is a
consequence of the law of large numbers. Also see that, as discussed
in Section~\ref{sec:geometry},
$\xi_\mathrm{min}\ge \underline{\xi}/2$.

\begin{figure}[t!]
  \centering 
  \includegraphics[height=6.5cm]{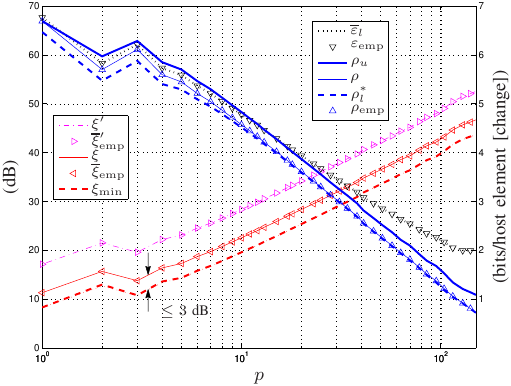}
  \caption{Performance of partitioned permutation coding for a
    quantized Gaussian host with $n=10^6$ as a function of the number
    of partitions, for the sequence of support partitionings
    $\{\mathcal{\widetilde{U}}_q^p\}$. Rate and efficiency units are
    on the right-hand side of the plot, whereas distortion units are
    on the left-hand side;
    $\rho_l^*=\max(\rho_l,\rho_l')$.} \label{fig:dwr-psnr-eff-rate}
\end{figure}

We verify next the results for partitioned permutation coding, using a
host $\mathbf{x}$ drawn from a Gaussian distribution with mean~$128$ and standard
deviation~$25$, and quantized to $\mathcal{V}=\{0,1,\ldots,2^b-1\}$
($b=8$). The sequence of support partitionings
$\{\mathcal{\widetilde{U}}_q^p\}$ discussed in
Section~\ref{sec:pract-select-part} is used to obtain the results in
Figure~\ref{fig:dwr-psnr-eff-rate}. The rate-distortion
bound~\eqref{eq:rd-function} is now key to the achieved performance,
whereas~\eqref{eq:fano} is too loose and not shown. Like in the
previous figure, we observe a close match between empirical
measurements for one-shot experiments and their corresponding
averages. This is also true for the match between the empirical
embedding efficiency and the lower bound on
$\overline{\varepsilon}$. It can be verified that, as discussed in
Section~\ref{sec:embedd-dist-contr} inequalities~\eqref{eq:xi_min_lb}
and~\eqref{eq:ineq_vareps} still hold. Finally, the most important
feature in Figure~\ref{fig:dwr-psnr-eff-rate} is the narrow nearly
constant gap $\rho_u-\rho_\text{emp}$ throughout, which is roughly the
same as the gap $\rho_u-\rho_l^*$ between the rate-distortion bounds.

\section{Concluding Remarks}
\label{sec:conclusions}
We have given a solution to the fundamental problem of asymptotically
optimum perfect universal steganography of finite memoryless sources
with a passive warden. We have shown that Slepian's Variant~I
permutation codes are central to this problem, and that they can be
efficiently implemented by means of adaptive arithmetic coding. This
reflects in practice the duality between perfect steganography and
lossless compression. We have also studied the embedding distortion of
permutation coding, and extended the problem above to include a distortion
constraint. The method that we have proposed for the constrained
scenario (partitioned permutation coding) performs close to an
unattainable upper bound on the rate-distortion function of the
problem.

In the same way that optimum lossless source coding of memoryless
signals is at the core of compression methods for real-world signals,
we expect that optimum perfect steganography of memoryless signals
will find its place at the core of future steganographic methods for
real-world host signals. Possible approaches include the use of
decorrelating integer-to-integer transforms, or decorrelation through
predictive techniques such as prediction by partial
matching~\cite{cleary84:_data_compr}, prior to the application of
permutation coding. Inevitably, steganography will no longer be
perfect when the memoryless assumption becomes only an approximation,
but the aforementioned approaches have the virtue of decoupling the
steganography problem from the decorrelation problem.  To conclude, we
should also note that the duality between steganography and source
coding also means that the distortion-constrained algorithm we have
presented may be of interest in the dual problem of lossless source
coding with side information at the decoder (i.e., distributed source
coding).

\appendix

\section{Appendix}

\label{sec:aver-over-symm}
In this appendix we obtain two expectations of matrix functions of a
random variable~$\mathit{\Pi}$ uniformly distributed over the ensemble
of all permutation matrices, i.e.,
$\Pr\{\mathit{\Pi}=\Pi_{\boldsymbol\sigma}\}=1/n!$ for all
${\boldsymbol\sigma}\in\mathcal{S}_n$. Firstly, we wish to calculate
\begin{eqnarray}\label{eq:av_perm_sn}
  \E\{\mathit{\Pi}\}&=&\frac{1}{n!}\sum_{\boldsymbol\sigma\in\mathcal{S}_n}\Pi_{\boldsymbol\sigma}.
\end{eqnarray}
In order to evaluate~\eqref{eq:av_perm_sn} consider the number of $n\times n$
permutation matrices which have a one at any given entry. This is
equivalent to fixing the corresponding row, and therefore there are
$(n-1)!$ possibilities for the remaining $n-1$ rows. Since this holds
for any entry, the summation on the right equals
$(n-1)!\,\mathbf{1}\mathbf{1}^t$, and so we have that
\begin{equation}
  \label{eq:sum_per}
  \E\{\mathit{\Pi}\}=\frac{1}{n}\,\mathbf{1}\mathbf{1}^t.
\end{equation}
Also see that, if we assume
$\Pr\{\mathit{\Pi}=\Pi_{\boldsymbol\sigma}\}=1/r$ for all
${\boldsymbol\sigma}\in\mathcal{S}_\mathbf{x}$, then the average
vector $\E\{\mathit{\Pi}\mathbf{x}\}$ can
be obtained using~\eqref{eq:sum_per} by observing that
\begin{equation}
\frac{1}{r}\sum_{\boldsymbol\sigma\in\mathcal{S}_\mathbf{x}}\Pi_{\boldsymbol\sigma}\mathbf{x}=
\frac{1}{n!}\sum_{\boldsymbol\sigma\in\mathcal{S}_n}\Pi_{\boldsymbol\sigma}\mathbf{x},
\label{eq:epix}
\end{equation}
which holds because the second sum contains the same
summands as the first one, but each repeated $h_1!h_2!\cdots
h_q!$ times.

Next we wish to compute
\begin{eqnarray}
  \label{eq:ex}
  \E\left\{\mathit{\Pi}\mathbf{x}\mathbf{x}^t\mathit{\Pi}^t\right\}&=&\frac{1}{n!}\sum_{\boldsymbol\sigma\in\mathcal{S}_n}\Pi_{\boldsymbol\sigma}\mathbf{x}\mathbf{x}^t\Pi_{\boldsymbol\sigma}^t,
\end{eqnarray}
for some arbitrary vector~$\mathbf{x}$. This average is a particular
case of the general formula given by Daniels
in~\cite[equation (4.9)]{daniels62:_processes}: %
\begin{equation*}
  \E\left\{\mathit{\Pi}\mathbf{x}\mathbf{x}^t\mathit{\Pi}^t\right\}=a\,\mathrm{I}+b\,
  \mathbf{1}\mathbf{1}^t, \label{eq:pxxp}
\end{equation*}
with $b= \left((\mathbf{x}^t\mathbf{1})^2-\|\mathbf{x}\|^2\right)/(n(n-1))$ and
  $a+b=(1/n)\|\mathbf{x}\|^2$.

Finally observe that average~\eqref{eq:ex} is the same if we replace
$n!$ and
$\mathcal{S}_n$ by $r$ and $\mathcal{S}_\mathbf{x}$, for the
same reason that makes expression~\eqref{eq:epix} true.

\section*{Acknowledgment}
The authors wish to thank the anonymous reviewers for their careful
reading of the manuscript and for their insightful comments, which
helped to clarify and improve the presentation of the materials.

\bibliographystyle{IEEEtran}

\begin{IEEEbiographynophoto}{Félix Balado} (M'03) graduated with an
  M.Eng. in Telecommunications Engineering from the University of Vigo
  (Spain) in 1996, and received a Ph.D. from the same institution in
  2003 for his work in digital data hiding. He is currently a lecturer
  in University College Dublin (Ireland). His research interests lie
  in the areas of signal processing, digital communications, and
  bioinformatics.
\end{IEEEbiographynophoto}

\begin{IEEEbiographynophoto}{David Haughton} received a Ph.D. from
  University College Dublin (Ireland) in 2014, for his work in the
  area of information embedding in DNA. He is currently a senior data
  scientist in the Centre for Applied Data Analytics Research in
  Dublin (Ireland).  His research background is in signal processing,
  bioinformatics and steganography, and his current research interest
  lies in big data analytics.
\end{IEEEbiographynophoto}

\vfill

\end{document}